\documentclass[12pt]{article}
\usepackage{datetime,graphicx,amssymb,amstext,amsmath,parskip,abstract,titlesec,adjustbox,multirow,bbold}
\usepackage[top=30mm, bottom=30mm, left=20mm, right=20mm]{geometry}
\usepackage{setspace}
\numberwithin{equation}{subsection}

\allowdisplaybreaks

\begin{document}

\begin{titlepage}

\begin{center}

\vspace*{0.5cm}
\LARGE
\textbf{Integrability of generalised type II defects in affine Toda field theory}

\vspace{1cm}

\large
\text{R. Bristow}

\emph{Department of Mathematical Sciences,\\
Durham University, Durham, U.K.\\
DH1 3LE}

\small
\text{\emph{Email:} rebecca.bristow@durham.ac.uk}

\vspace{2cm}

\end{center}
\normalsize
The Liouville integrability of the generalised type II defects is investigated. Full integrability is not considered, only the existence of an infinite number of conserved quantities associated with a system containing a defect. For defects in affine Toda field theories (ATFTs) it is shown that momentum conservation is very likely to be a necessary condition for integrability. The defect Lax matrices which guarantee zero curvature, and so an infinite number of conserved quantities, are calculated for the momentum conserving Tzitz\'{e}ica defect and the momentum conserving $D_4$ ATFT defect. Some additional calculations pertaining to the $D_4$ defect are also carried out to find a more complete set of defect potentials than has appeared previously.
\end{titlepage}

\section{Introduction}

In this paper we consider the integrability of the defects in affine Toda field theories (ATFTs) found in \cite{bb17}. Since (some of) the interest in integrable systems is due to their ability to model physical phenomena whilst remaining exactly solvable it is important to be able to incorporate common physical occurences without destroying the integrability of the system. A defect is some discontinuity in physical media or fields in a mathematical model, and we will check whether incorporating a discontinuity into an integrable model can be achieved without destroying its integrability. We follow the classical Lagrangian picture of defects introduced in \cite{bcz04a} and further studied in \cite{bcz04b,cz07,cz09a,cz09b,rob14a,rob15,bb17}. In this approach for a defect at $x=0$ there is a field vector $u$ defined in the region $x\leq 0$ and a field vector $v$ defined in the region $x\geq 0$, with both fields obeying the same bulk theory. There may also be some additional degrees of freedom appearing only at the defect, which are referred to as auxiliary fields. The Lagrangian density of the whole system then contains the Lagrangians of the bulk theories, restricted to the appropriate regions, and a defect term coupling the two sets of bulk fields and any auxiliary fields at $x=0$. This defect term consists of a ``kinetic" part, containing time derivatives of the fields, and a defect potential. Using this Lagrangian density in the Euler-Lagrange equations will yield the bulk equations of motion for $u$ restricted to $x\leq 0$, the bulk equations of motion for $v$ restricted to $x\geq 0$, and some equations of motion coupling the bulk fields $u$ and $v$ to each other and to the auxiliary fields (if they are present), evaluated at $x=0$.

For classical $1+1$ dimensional field theories Liouville integrability is defined as possessing an infinite number of independent conserved quantities in Poisson involution. Such a system is in principle solvable \cite{ft86,bbt03}. Solitons are a particular set of solutions which are a feature of integrable systems and appear as stable, localised field configurations. There are many physical examples of integrable systems and solitons, for just a few of these see \cite{scm73,for90}. One method of proving the integrability of a system is using the method of Lax pairs and r-matrices first introduced in \cite{lax68}.

The Lax pair is a pair of matrices $a_0(t,x,\lambda)$ and $a_1(t,x,\lambda)$ such that for a vector field $\Psi(t,x)$
\begin{align}
\frac{\mathrm{d} \Psi(t,x)}{\mathrm{d} t} =& -a_0(t,x,\lambda)\Psi(t,x) \label{eq:Laxt} \\
\frac{\mathrm{d} \Psi(t,x)}{\mathrm{d} x} =& -a_1(t,x,\lambda)\Psi(t,x) \label{eq:Laxx}
\end{align}
where $\lambda$ is the spectral parameter. These Lax matrices may be used to transport the vector $\Psi$ as
\begin{align}
\Psi(t_2,x,\lambda) =& Pe^{-\int_{t_1}^{t_2} \mathrm{d}t' a_0(t',x,\lambda) } \Psi(t_1,x,\lambda) \label{eq:transportt} \\
\Psi(t,x_2,\lambda) =& Pe^{-\int_{x_1}^{x_2} \mathrm{d}x' a_1(t,x',\lambda) } \Psi(t,x_1,\lambda) \label{eq:transportx}
\end{align}
where $P$ denotes path ordering. The transport matrices themselves are also solutions to eqs.\eqref{eq:Laxt}, \eqref{eq:Laxx} respectively. By either requiring that the overdetermined system of equations in eqs.\eqref{eq:Laxt}, \eqref{eq:Laxx} are consistent, or that the transport as given in eqs.\eqref{eq:transportt}, \eqref{eq:transportx} is path independent, we find the zero curvature condition to be
\begin{align}
a_{1,t} -a_{0,x} +[a_0,a_1] =& 0 \label{eq:zcc}
.\end{align}
This must be satisfied by the Lax pair if we are to generate an infinite number of conserved quantities. The gauge transformation
\begin{align}
a_0 &\rightarrow \tilde{a}_0=-G_tG^{-1} +Ga_0G^{-1} \label{eq:a0GT} \\
a_1 &\rightarrow \tilde{a}_1=-G_xG^{-1} +Ga_1G^{-1} \label{eq:a1GT}
\end{align}
leaves the zero curvature condition unchanged.

The system in the bulk is some field $u$ (or $v$) which is governed by an equation of motion. If a pair of matrices which are dependent on $u$ and the spectral parameter $\lambda$ satisfy eq.\eqref{eq:zcc} if and only if $u$ satisfies the equations of motion of the system then we have a Lax pair of the system. This Lax pair may then always be used to generate an infinite number of conserved quantities involving the field $u$, and thus being conserved quantities of the integrable system. To generate these conserved quantities the Lax pair is used to give the monodromy matrix, which transports $\Psi$ between $x\rightarrow -\infty$ and $x\rightarrow\infty$. The trace of this matrix is equal when evaluated at different times, and it is possible to expand this in terms of the spectral parameter $\lambda$ and equate powers of $\lambda$ to give an infinite number of conserved quantities. If the system is integrable it is then possible to construct a related r-matrix, which will ensure that these conserved quantities Poisson commute.

The integrable field theories which we will be considering here are the ATFTs. They began life as a description of a one-dimensional lattice of particles with nearest-neighbour interactions, which was shown to be integrable with soliton solutions \cite{tod70}. The potential of this system contained terms of the form $e^{u_{i-1}-u_i}$, where $u_i$ is the position of particle $i$, and in \cite{bog76} these potential terms were generalised to depend on the simple roots of any Lie algebra. The ``affine" refers to the fact that the potential is written in terms of the simple roots and the lowest weight root, as the addition of the lowest weight root to a Dynkin diagram gives an affine Dynkin diagram. In \cite{mik79} the Toda lattice is taken to a two-dimensional field theory for the $A_r$ and Tzitz\'{e}ica cases. All ATFTs are given in \cite{mop81} and their conserved quantities are investigated. These were first shown to have zero curvature (and so an infinite number of conserved quantities) \cite{mop81,wil81} and later shown to be integrable \cite{ot85,ot86} using the method of the Lax pair and r-matrix. All ATFTs have solitons as solutions \cite{hol92,mm93,mcg94a}. As well as being integrable solitons (stable by virtue of a cancellation of nonlinear and dispersive forces) these solitons are also topological (stable due to possessing some topological charge, in this case the difference between the field as $x\rightarrow\pm\infty$).

An ATFT is described by the Lagrangian density
\begin{align}
\mathcal{L}^u = \frac{1}{2}u_{i,t}u_{i,t} -\frac{1}{2}u_{i,x}u_{i,x} -U
\quad\quad\quad\quad
U = \frac{m^2}{\beta^2} \sum_{i=0}^r n_i e^{\beta(\alpha_i)_ju_j} \label{eq:Lbulk}
\end{align}
where $\alpha_i$ ($i=1,\dots,r$) are the simple root vectors of a Lie algebra, $n_i$ ($i=1,\dots,r$) are a set of integers characteristic of each algebra, $n_0=1$ and $\alpha_0=-\sum_{i=1}^r n_i \alpha_i$ gives the lowest weight root which corresponds to the extra node on an affine Dynkin diagram. $m$ is the mass constant, $\beta$ is the coupling constant and in the classical case we can rescale the field $u$ and the variables $t$ and $x$ to set $m=\beta=1$. Taking this expression with $v$ instead of $u$ gives the Lagrangian density $\mathcal{L}^v$ and the potential $V$ which will goven the behaviour of the field to the right of the defect. The vector $u=(u_1,\dots,u_r)^T$ lies in the space spanned by the simple root vectors and the fields $\{u_i\}$ are the projections of $u$ onto the basis of this vector space.

Because the simple roots are defined only up to their inner products with other simple roots the potential based on the set of roots $\{\alpha_i\}$ and the potential based on the set of roots $\{Q\alpha_i\}$, where $Q$ is some orthogonal transformation, describe the same ATFT. Because the kinetic part of the bulk Lagrangian is invariant under orthogonal transformations of the fields the ATFTs based on the roots $\{\alpha_i\}$ can be obtained by taking $u\rightarrow Q u$ in the ATFT based on the roots $\{Q\alpha_i\}$. In a similar manner we can take the ATFT based on $\{c\alpha_i\}$, where $c$ is a constant, and, with $u\rightarrow c^{-1}u$ and a rescaling of the coordinates $t$ and $x$ such that $\partial_{t,x}\rightarrow c\partial_{t,x}$, return to the ATFT based on the roots $\{\alpha_i\}$. Therefore our precise choice of root vectors is unimportant, and they can be set to be as simple as possible.

This potential has multiple vacua occurring at $2\pi i$ multiples of weights of the Lie algebra whose simple roots the potential is based on, so if the field $u$ is complex then we can have soliton solutions to the ATFT equations of motion which interpolate between different vacua as $x\rightarrow\pm\infty$. Such soliton solutions have been found for all ATFTs \cite{hol92,mm93,mcg94a,hal94}.

For an ATFT the Lax pair is
\begin{align}
a_0 =& \frac{1}{2}\left( u_x.H +\frac{1}{\sqrt{2}}\sum_{i=0}^r \sqrt{n_i} |\alpha_i| e^{\frac{1}{2}\alpha_i.u}\left( \lambda E_{\alpha_i} -\frac{1}{\lambda} E_{-\alpha_i} \right) \right) \label{eq:ATFTa0} \\
a_1 =& \frac{1}{2}\left( u_t.H +\frac{1}{\sqrt{2}}\sum_{i=0}^r \sqrt{n_i} |\alpha_i| e^{\frac{1}{2}\alpha_i.u}\left( \lambda E_{\alpha_i} +\frac{1}{\lambda} E_{-\alpha_i} \right) \right) \label{eq:ATFTa1}
\end{align}
\cite{mop81} where $H$ are the Cartan generators and $E_{\alpha_i}$ is the generator associated with the root $\alpha_i$. While we are using the affine simple roots we are still using the non-affine, finite dimensional generators which obey the commutation relations
\begin{align}
[H_j,E_{\alpha}] =& (\alpha)_jE_{\alpha} \label{eq:HEE} \\
E_{\alpha} =& E_{-\alpha}^{\dagger} \label{eq:EE} \\
[E_{\alpha},E_{-\alpha}] =& \frac{2}{|\alpha|^2}(\alpha)_jH_j \label{eq:EEH} \\
[E_{\alpha},E_{\beta}] =& n_{\alpha\beta}E_{\alpha+\beta} & &\mathrm{if} \quad \alpha+\beta \in \mathrm{roots} \label{eq:EEE} \\
[E_{\alpha},E_{\beta}] =& 0 & &\mathrm{if} \quad \alpha+\beta \notin \mathrm{roots},0 \label{eq:EE0}
.\end{align}
Here subscripts are used to identify the different generator matrices and roots. A subscript outside a bracket denotes a component of the bracketed vector. From eq.\eqref{eq:HEE} we see that the Cartan generator $H_i$ is associated with the projections of the roots onto the basis vector $e_i$, hence each Cartan generator is associated with one of the orthonormal basis vectors of the root space and we can take $u.H=u_iH_i$. Using this Lax pair in eq.\eqref{eq:zcc}, along with these commutation relations, we can check that it is satisfied provided that the equations of motion of the ATFT (given by the Lagrangian density in eq.\eqref{eq:Lbulk}) are satisfied, and so our bulk theories have zero curvature.

Some of the earliest studies of defects were in quantum integrable field theories, for example in a free fermion theory \cite{dms94a,dms94b} and in sine-Gordon theory \cite{kl99}, and here it was shown that integrable defects must be purely reflecting or transmitting. From the fact that quantum defects must be purely transmitting (a purely reflecting defect is simply a boundary, as investigated in \cite{bcdr95}) came the idea that momentum conservation may be important in the classical case.

In \cite{bcz04a} it was found that for a defect in sine-Gordon theory certain defect equations ensured that momentum was conserved. The conservation of energy and some higher spin charges was also checked for these momentum conserving defects. These defects which couple the bulk fields $u$ and $v$, but have no auxiliary fields, are referred to as type I defects, and were generalised to give momentum conserving defects in $A_r$ ATFTs. However it was also proved that momentum conserving defects of the particular form found in \cite{bcz04a,bcz04b} could never appear in an ATFT based on a Lie algebra other than $A_r$. In \cite{cz09b} the momentum conserving defects first found in \cite{bcz04a} were modified by the addition of a degree of freedom at the defect, allowing a momentum conserving defect in the Tzitz\'{e}ica model (previously excluded due to not being based on the roots of $A_r$) to be found. These defects with auxiliary fields are referred to as type II defects. This idea of extra fields at the defect, and the fact that one ATFT can be folded to a different ATFT using certain symmetries of the Dynkin diagram \cite{ot83a,ot83b}, was used in \cite{rob14a} to fold existing $A_r$ ATFT defects to new $C_r$ ATFT defects. These type II defects were generalised in \cite{bb17} and momentum conserving defects were found in the $B_r$ and $D_r$ ATFTs. Some investigation into defects in non-relativistic theories such as the nonlinear Schr\"{o}dinger equation and the Kortweg-de Vries equation have also been made \cite{cz06,cp16}.

That it is possible for a system which explicitly breaks time translation invariance to have conserved momentum is very interesting, and it is hoped that the restrictions arising from momentum conservation are sufficient to ensure the integrability of the system. There are already some strong indications that this is the case. All of the above defects have soliton solutions in which a soliton passes from one bulk theory to the other, experiencing a delay and sometimes a change in topological charge. An interesting consequence of requiring momentum conservation is that the defect equations can always be modified in such a way that they give a B\"{a}cklund transformation (some first order differential equations coupling the solutions to two sets of uncoupled higher order differential equations \cite{miu76}) for the bulk theories \cite{bcz04a,bcz04b,cz09b,bb17}.

The type I defects have already been shown to possess an infinite number of conserved quantities, and this along with the soliton solutions indicates that they are likely integrable \cite{bcz04b,cz07,cz09a}. However, the integrability of these particular defects has not been proven as they are given in a Lagrangian rather than a Hamiltonian form, meaning that the Poisson brackets and r-matrix required to prove that the charges are in involution are difficult to write down. A type II defect matrix for the Tzitz\'{e}ica model is found in \cite{aagz11} and the system is shown to have an infinite number of conserved quantities. A Hamiltonian set-up in which the Lax and r-matrix equations are immediately assumed to be satisfied by some matrix associated with the defect is investigated in \cite{ad12a,ad12b,doi15b,doi16} for defects in the nonlinear Schr\"{o}dinger equation, sine-Gordon and ATFTs. While these defects are integrable they do not necessarily describe the same systems as the momentum conserving defects found in the Lagrangian set-up. Some attempt to reconcile this Hamiltonian approach and the Lagrangian approach to defects is made in \cite{cau15,ck15}. The type I and type II Lagrangians are rewritten as Hamiltonians with second class constraints in \cite{cz09b}.

In this paper we will not attempt to prove the integrability of a system with a defect, only that it posesses an infinite number of conserved quantities. We will achieve this using the method of zero curvature and Lax pairs developed for the Kortweg-de Vries equation in \cite{lax68} and modified to apply to a system with a type I defect in \cite{bcz04b}. We first give a recap of momentum conserving defects, giving the generalised type II defects found in \cite{bb17} in section \ref{sec:mcd}, with these momentum conserving defects in ATFTs given in section \ref{sec:mcd atft}. Section \ref{sec:mcd tzitz} gives the Tzitz\'{e}ica defect found in \cite{cz09b} and section \ref{sec:mcd d4} gives some new, more complete results for a momentum conserving defect in the $D_4$ ATFT. In section \ref{sec:zcc} we show the derivation of the zero curvature condition for a defect which appeared in \cite{bcz04b}.  In section \ref{sec:zcc atft} we find that for a defect in an ATFT momentum conservation can be shown to be a necessary condition for zero curvature of the system, and are able to find some possible restrictions on the exact form an integrable defect may take. However, we are unable to prove that the zero curvature condition for these defects can be satisfied. Finally in sections \ref{sec:zcc tzitz} and \ref{sec:zcc d4} we consider the zero curvature of two specific defects, those in the Tzitz\'{e}ica model and the $D_4$ ATFT. The defect matrix for the Tzitz\'{e}ica model has been found previously in \cite{aagz11}. Both are shown to satisfy the zero curvature condition, and so have an infinite number of conserved quantities.

\section{Momentum conserving defects}

We will now present the results on generalised type II momentum conserving defects which appear in \cite{bb17}, with the defect Lagrangian and potential for momentum conserving defects in any ATFT, the Tzitz\'{e}ica model and the $D_4$ ATFTs.

\subsection{Generalised momentum conserving type II defects} \label{sec:mcd}

Components of the bulk fields $u$ and $v$ are denoted as $u_1,u_2,\dots$, $v_1,v_2,\dots$ and because we assume that $u$ and $v$ describe two copies of the same bulk theory the number of components of $u$ and $v$ are equal. The auxiliary fields at the defect are contained in the field vector $\lambda$, with components denoted by $\lambda_1,\lambda_2,\dots$. There may be any number of components of the auxiliary field vector. Note that this field vector $\lambda$ is not the spectral parameter; we specify whether $\lambda$ is the auxiliary field vector or the spectral parameter whenever it appears in this paper. The Lagrangian description of the theory in the presence of a defect at $x=0$ is given in terms of a density
\begin{align}
\mathcal{L} = \Theta(-x)\mathcal{L}^{u} +\Theta(x)\mathcal{L}^{v} +\delta(x)\mathcal{L}^D, \label{eq:Loverall}
\end{align}
where the bulk Lagrangian densities
\begin{align}
\mathcal{L}^{u}&=\frac{1}{2}(u_{i,t}u_{i,t}-u_{i,x}u_{i,x})-U(u)\label{eq:bulkLu} \\
\mathcal{L}^{v}&=\frac{1}{2}(v_{i,t}v_{i,t}-v_{i,x}v_{i,x})-V(v)\label{eq:bulkLv}
\end{align}
govern the behaviour of the bulk fields $u$ and $v$. Subscripts of $t$ and $x$ denote partial differentiation with respect to that variable and are separated from subscripts of indices by a comma. Einstein sum notation is used throughout. The two bulk theories are coupled at $x=0$ via the  defect Lagrangian $\mathcal{L}^D$ which depends on $u$, $v$ and $\lambda$. Note that this form of defect is not restricted to the ATFTs. This Lagrangian set-up was pioneered in \cite{bcz04a}.

Motivated by the form of the type I defects appearing in \cite{bcz04a,bcz04b,cz07,cz09a} and the type II defects appearing in \cite{cz09b,rob14a} the generalised type II defect Lagrangian density was taken to be
\begin{align}
\mathcal{L}^D =& \frac{1}{2}u_iA_{ij}u_{j,t} +\frac{1}{2}v_iB_{ij}v_{j,t} +u_iC_{ij}v_{j,t} +\frac{1}{2}\lambda_iW_{ij}\lambda_{j,t} +\lambda_iX_{ij}u_{j,t} +\lambda_iY_{ij}v_{j,t} -F(u,v,\lambda) \label{eq:defectLgeneral}
\end{align}
where $A$, $B$, $C$, $W$, $X$ and $Y$ are arbitrary, constant, real coupling matrices.

Because the auxiliary field vector $\lambda$ does not appear in the bulk Lagrangians the behaviour of the system is not altered under the redefinition of the auxiliary fields $\lambda_i \rightarrow \alpha_{ij}u_j +\beta_{ij}v_j +\gamma_{ij}\lambda_j$. $\alpha$ and $\beta$ are any matrices and $\gamma$ is an invertible matrix to ensure the degrees of freedom associated to the auxiliary fields are not removed. The bulk fields can also be transformed as $u_i \rightarrow Q_{ij}u_j$, $v_i \rightarrow Q'_{ij}v_j$ without changing the general form of the bulk and defect Lagrangians provided $Q$ and $Q'$ are both orthogonal.

Energy and momentum were the only conserved charges investigated in \cite{bb17}, with momentum conservation proving to be particularly restrictive. Provided $\{u_i\},\{v_i\}\rightarrow \text{constant}$ as $x\rightarrow \pm\infty$ and $U$ and $V$ have no local minima the energy of the system in the bulk differentiated with respect to $t$ is
\begin{align}
\frac{\mathrm{d}E}{\mathrm{d}t} =& \left.\left(u_{i,x}u_{i,t} -v_{i,x}v_{i,t}\right)\right|_{x=0} \label{eq:Et}
.\end{align}
Using the defect conditions arising from eq.\eqref{eq:defectLgeneral} to remove the $x$ derivatives we find that eq.\eqref{eq:Et} may be rewritten as
\begin{align}
\frac{\mathrm{d}E}{\mathrm{d}t} = -\frac{\mathrm{d}F}{\mathrm{d}t} \label{eq:EtFt}
.\end{align}
Therefore $E+F$ is the conserved energy-like quantity, where $E$ is the bulk energy and $F$ is the defect potential. Since the defect breaks manifest translation invariance and so the system is no longer obviously momentum conserving it was expected that requiring conservation of momentum would be far more restrictive than requiring conservation of energy. The momentum of the system in the bulk differentiated with respect to $t$ is
\begin{align}
\frac{\mathrm{d}P}{\mathrm{d}t} = \left.\left(\frac{1}{2}\left( u_{i,t}u_{i,t} +u_{i,x}u_{i,x} -v_{i,t}v_{i,t} -v_{i,x}v_{i,x} \right) -U +V \right)\right|_{x=0} \label{eq:Pt}
.\end{align}
For the system to be momentum conserving we must be able to use the defect equations arising from eq.\eqref{eq:defectLgeneral} to rewrite the right hand side of this equation as a total time derivate. This places certain constraints on both the coupling matrices and the defect potential.

By using this freedom to make field redefinitions and by applying the constraints arising when the system is taken to conserve momentum this defect Lagrangian was rewritten as
\begin{align}
\mathcal{L}^D =& \frac{1}{2}u^{(1)}_{i}A_{ij}u^{(1)}_{j,t} +\frac{1}{2}v^{(1)}_{i}A_{ij}v^{(1)}_{j,t} +u^{(1)}_{i}\left(\mathbb{1}-A\right)_{ij}v^{(1)}_{j,t} \nonumber \\
&+u^{(2)}_{i}v^{(2)}_{i,t} +2\mu^{(2)}_{i}\left(u^{(2)}_{i,t} -v^{(2)}_{i,t}\right) +\frac{1}{2}\xi_{i}W_{ij}\xi_{j,t} -F \label{eq:finaldefectL}
.\end{align}
The components of $\lambda$ which (after field redefinitions) coupled to no bulk fields, only other auxiliary fields, are contained in the vector $\xi$, with the coupling matrix $W$ given by
\begin{align}
W=\left(\begin{matrix}
0 & 1 & \hdots & 0 & 0 \\
-1 & 0 & \hdots & 0 & 0 \\
\vdots & \vdots & \ddots & \vdots & \vdots \\
0 & 0 & \hdots & 0 & 1 \\
0 & 0 & \hdots & -1 & 0 \\
\end{matrix}\right) \label{eq:W}
.\end{align}
The remaining auxiliary fields, which do couple to the bulk fields, are contained in the vector $\mu^{(2)}$. The form of the couplings of the bulk fields and these auxiliary fields are partly determined by requiring momentum conservation and partly by our choice of field redefinitions, intended to simplify the various couplings as far as possible. The vector $\xi$ contains $m$ components, the vector $\mu^{(2)}$ contains $n$ components and the bulk vectors $u$ and $v$ have $r$ components. Of the components of $u$ and $v$, $n$ couple to some auxiliary field (with every component of $\mu^{(2)}$ coupling to a different pair of bulk fields) and $n-r$ do not. The bulk fields which do not couple to any auxiliary fields are contained in the vectors $u^{(1)}$ and $v^{(1)}$, so labelled because they couple like the fields in a type I defect. The coupling matrix $A$ may be any antisymmetric matrix. The bulk fields which do couple to auxiliary fields are contained in the vectors $u^{(2)}$ and $v^{(2)}$, with the labelling indicating coupling similar to that in a type II defect. The bulk fields may be split between the $(1)$ and $(2)$ vectors, and the auxiliary fields between the $\mu^{(2)}$ and $\xi$ vectors, in any way (provided $\mu^{(2)}$, $u^{(2)}$ and $v^{(2)}$ all have the same number of components). For the full calculation taking the general defect Lagrangian in eq.\eqref{eq:defectLgeneral} to the momentum conserving defect Lagrangian in eq.\eqref{eq:finaldefectL} see \cite{bb17}.

It was shown in \cite{bb17} that every momentum conserving defect must be related to this form of defect Lagrangian by a field redefinition of the auxiliary fields or an orthogonal transformation of the bulk fields. The particular choices of field redefinitions made to reach this form of the Lagrangian were intended to simplify the couplings as far as possible.

That the defect Lagrangian is in the form eq.\eqref{eq:finaldefectL} is a necessary but not yet sufficient condition for the defect to be momentum conserving. In addition to the ``kinetic" part of the defect Lagrangian being in the form given in eq.\eqref{eq:finaldefectL} the defect potential must be given by $F=D+\bar{D}$, where the dependencies of $D$ and $\bar{D}$ are
\begin{align}
D =& D\left(p^{(1)}+Aq^{(1)},p^{(2)}-\mu^{(2)},q^{(2)},\xi\right) \label{eq:Ddependencies} \\
\bar{D} =& \bar{D}\left(q^{(1)},q^{(2)},\mu^{(2)},\xi\right) \label{eq:Dbardependencies}
\end{align}
and they satisfy the momentum conservation condition
\begin{align}
2(U-V) =& D_{p^{(1)}_i}\bar{D}_{q^{(1)}_i} +D_{q^{(2)}_i}\bar{D}_{\mu^{(2)}_i} -D_{\mu^{(2)}_i}\bar{D}_{q^{(2)}_i} -4D_{\xi_{i}}W_{ij}\bar{D}_{\xi_{j}} \label{eq:mcc}
.\end{align}
The new field vectors $p$ and $q$ are given by $p_i=\frac{1}{2}\left(u_i+v_i\right)$, $q_i=\frac{1}{2}\left(u_i-v_i\right)$, with the components split between $p^{(1)}$, $q^{(1)}$ and $p^{(2)}$, $q^{(2)}$ in exactly the same way as the $u$ and $v$ field vectors split into $u^{(1)}$, $v^{(1)}$ and $u^{(2)}$, $v^{(2)}$. The total conserved energy and momentum of the system are $E+D+\bar{D}$ and $P+D-\bar{D}$, where $E$ and $P$ are the bulk energy and momentum.

A redefinition
$
\mu^{(2)}_i\rightarrow\mu^{(2)}_i+f\left(q^{(2)}\right)_{q^{(2)}_i}
$
does not alter the defect Lagrangian in eq.\eqref{eq:finaldefectL} as it only introduces a total $t$ derivative. Redefinitions of the bulk fields which are the orthogonal transformations $u^{(1)}\rightarrow Qu^{(1)}$ and $v^{(1)}\rightarrow Q^Tu^{(1)}$, or the orthogonal transformations $u^{(2)}\rightarrow Q'u^{(2)}$, $v^{(2)}\rightarrow Q'v^{(2)}$ and $\mu^{(2)}\rightarrow Q'^T\mu^{(2)}$, or the shifts $u\rightarrow u+c$, $v\rightarrow v+d$ (where $Q$ and $Q'$ are any orthogonal matrices and $c$ and $d$ are any constants) alter neither the bulk nor the defect Lagrangian. This means that once $D$ and $\bar{D}$ satisfying the momentum conservation condition have been found these field redefinitions can be used to give a family of different defect potentials satisfying the same momentum conservation condition. 

The equations of motion at the defect, with the defect Lagrangian given in eq.\eqref{eq:finaldefectL} with $F=D+\bar{D}$ and written in terms of $p_i=\frac{1}{2}(u_i+v_i)$, $q_i=\frac{1}{2}(u_i-v_i)$, are
\begin{align}
p^{(1)}_{i,x} =& 
p^{(1)}_{i,t}
+2A_{ij}q^{(1)}_{j,t}
-\frac{1}{2}D_{q^{(1)}_i} -\frac{1}{2}\bar{D}_{q^{(1)}_i} \label{eq:defecteom1} \\
q^{(1)}_{i,x} =&
-q^{(1)}_{i,t}
-\frac{1}{2}D_{p^{(1)}_i} \label{eq:defecteom2} \\
p^{(2)}_{i,x} =&
p^{(2)}_{i,t}
-2\mu^{(2)}_{i,t}
-\frac{1}{2}D_{q^{(2)}_i} -\frac{1}{2}\bar{D}_{q^{(2)}_i} \label{eq:defecteom3} \\
q^{(2)}_{i,x} =&
-q^{(2)}_{i,t}
-\frac{1}{2}D_{p^{(2)}_i} \label{eq:defecteom4} \\
0 =&
q^{(2)}_{i,t}
-\frac{1}{4}D_{\mu^{(2)}_i} -\frac{1}{4}\bar{D}_{\mu^{(2)}_i} \label{eq:defecteom5} \\
0 =& \xi_{i,t}
+W_{ij}D_{\xi_j} +W_{ij}\bar{D}_{\xi_j} \label{eq:defecteom6}
.\end{align}

Requiring momentum conservation is evidently very restrictive on the form the couplings at the defect and the defect potential may take. In the type I case requiring the defect to have zero curvature showed that the restrictions on the defect which ensured energy and momentum conservation were necessary and sufficient to ensure the existence of an infinite number of conserved charges \cite{bcz04b,cz09a}. We aim to show the same for the defects described in this section.

\subsection{Momentum conserving defects in ATFTs} \label{sec:mcd atft}

Recall that for the defect in eq.\eqref{eq:finaldefectL} we were required to split the bulk field components between vectors $u^{(1)}$ and $u^{(2)}$. For an ATFT $u$ lives in the root space of the underlying Lie algebra, so we can divide this vector space into the 1-space, with the projection of $u$ on to this space being $u^{(1)}$, and the 2-space, with the projection of $u$ onto this space being $u^{(2)}$. The 1-space and 2-space are orthogonal and sum to the root space. We then have $(\alpha_i)^{(1)}$ as the projection of the simple root $\alpha_i$ onto the 1-space and $(\alpha_i)^{(2)}$ as its projection onto the 2-space.

We can take the momentum conservation condition in eq.\eqref{eq:mcc} and use the ATFT potential in eq.\eqref{eq:Lbulk} (dependent on $u$ for $U$ and on $v$ for $V$), along with the dependencies of $D$ and $\bar{D}$ given in eqs.\eqref{eq:Ddependencies}, \eqref{eq:Dbardependencies} to see that they must take the form
\begin{align}
D =& \sigma\sum_{i=0}^n x_i\left(q^{(2)},\xi\right) e^{ \left(\alpha_i\right)^{(1)}_{j}\left(p^{(1)}_{j}+A_{jk}q^{(1)}_{k}\right) +(\alpha_i)^{(2)}_{j}\left(p^{(2)}_{j}-\mu^{(2)}_j\right)} \label{eq:Dgeneral} \\
\bar{D} =& \frac{1}{\sigma}\sum_{i=0}^n y_i\left(q^{(1)},q^{(2)},\xi\right) e^{ -(\alpha_i)^{(1)}_{j} A_{jk}q^{(1)}_{k} +(\alpha_i)^{(2)}_{j}\mu^{(2)}_j } \label{eq:Dbargeneral}
.\end{align}
This arises from considering the exponentials of the field $p$ which appear in $U-V$ and the dependencies of $D$ and $\bar{D}$. The defect parameter $\sigma$ is a free constant and appears because all terms in eq.\eqref{eq:mcc} are of the form $D\bar{D}$. $x_i$ and $y_i$ are functions yet to be determined. A major difficulty in finding $D$ and $\bar{D}$ which satisfied the momentum conservation condition was that there is no systematic way of determining how the root space should split into the 1-space and the 2-space. Trial and error was used to give all the results in \cite{bb17}.

Using eqs.\eqref{eq:Dgeneral}, \eqref{eq:Dbargeneral} in the momentum conservation condition and equating powers of $p$ we have
\begin{align}
2n_i\left( e^{(\alpha_i)_kq_k} -e^{-(\alpha_i)_kq_k} \right) =
\sum_{j=0}^r\Big( &
x_i(\alpha_i)_ky_{j,q_k}
+x_iy_j(\alpha_i)^{(1)}_{k}A_{kl}(\alpha_j)^{(1)}_{l}
+x_{i,q^{(2)}_{k}}(\alpha_j)^{(2)}_{k}y_j \nonumber \\
&-4x_{i,\xi_k}W_{kl}y_{j,\xi_l} \Big)e^{(\alpha_i-\alpha_j)^{(1)}_{k}A_{kl}q^{(1)}_{l}-(\alpha_i-\alpha_j)^{(2)}_{k}\mu_k} \label{eq:xymcc}
\end{align}
for $i=0,\dots,r$ as the momentum conservation conditions. We will give the solutions to these conditions for the Tzitz\'{e}ica and $D_4$ ATFT cases.

\subsection{Momentum conserving defects in the Tzitz\'{e}ica model} \label{sec:mcd tzitz}

This momentum conserving type II Tzitz\'{e}ica defect first appeared in \cite{cz09b}.

The Tzitz\'{e}ica potential is given by eq.\eqref{eq:Lbulk} with simple (and lowest weight) roots
\begin{align}
\alpha_0 =& -2 & \alpha_1 =& 1 \label{eq:Tzitzroots}
\end{align}
and marks
\begin{align}
n_0 =& 1 & n_1 =& 2 \label{eq:Tzitzmarks}
.\end{align}

The bulk fields are evidently scalar, and from \cite{cz09b} we know that there will be a scalar auxiliary field. The defect Lagrangian is
\begin{align}
\mathcal{L}^D =& uv_{t} +2\mu\left(u_{t} -v_{t}\right) -D -\bar{D} \label{eq:Tzitzdefect}
\end{align}
and for this to be momentum conserving $D(p-\mu,q)$ and $\bar{D}(q,\mu)$ (with $p=\frac{1}{2}(u+v)$, $q=\frac{1}{2}(u-v)$) must satisfy the momentum conservation condition
\begin{align}
2\left(e^{-2(p+q)}-e^{-2(p-q)}+2e^{p+q}-2e^{p-q}\right) =& D_q\bar{D}_{\mu} -D_{\mu}\bar{D}_q \label{eq:Tzitmcc}
.\end{align}
Because only $D$ is dependent on $p$ and the right hand side must be overall independent of $\mu$ we can write
\begin{align}
D =& \sigma\left( x_0(q) e^{-2p+2\mu} +x_1(q) e^{p-\mu} \right) \label{eq:TzitD} \\
\bar{D} =& \frac{1}{\sigma}\left( y_0(q) e^{-2\mu} +y_1(q) e^{\mu} \right) \label{eq:TzitDbar}
.\end{align}
At the end of section \ref{sec:mcd} we noted that the redefinition $\mu\rightarrow\mu+f(q)$ of the auxiliary field, where $f$ is any function, does not change the kinetic part of the defect Lagrangian and so can be used to give a family of $D$ and $\bar{D}$ satisfying the same momentum conservation condition. In order to simplify the differential equations to be solved we will use the field redefinition $\mu\rightarrow\mu-\frac{1}{2}\ln{x_1}$ to set $x_1=1$. The other coefficients are currently arbitrary, so can be redefined to include this.

Using these choices for the Tzitz\'{e}ica simple roots, marks and choice of 1-space and 2-space (i.e. no 1-space and a one dimensional 2-space) in the set of differential equations which are the momentum conservation conditions for a general ATFT in eq.\eqref{eq:xymcc} we can write down a set of differential equations to be solved for $x_{0,1}$ and $y_{0,1}$. This set of four differential equations which form the momentum conservation condition are then solved by
\begin{align}
x_0 =& \frac{1}{2c}(e^{q}+e^{-q})^2 & y_0 =& c \nonumber \\
x_1 =& 1 & y_1 =& 4(e^q+e^{-q}) \label{eq:Tzitzxy}
\end{align}
where $c$ is a constant. We now have a specific solution,
\begin{align}
D =& \sigma\left(\frac{1}{2c}(e^{q}+e^{-q})^2e^{-2p+2\mu} +e^{p-\mu}\right) \label{eq:TzitD2} \\
\bar{D} =& \frac{1}{\sigma}\left(ce^{-2\mu} +4(e^q+e^{-q})e^{\mu}\right) \label{eq:TzitDbar2}
.\end{align}
We can choose to take $\mu\rightarrow\mu+\frac{1}{3}\ln{c}$ and redefine the defect parameter as $\sigma\rightarrow c^{\frac{1}{3}}\sigma$. This removes all instances of the constant $c$. To introduce as much freedom as is possible we then make the field redefinition $\mu\rightarrow\mu+f(q)$, giving
\begin{align}
D =& \sigma\left( \frac{1}{2}(e^{q}+e^{-q})^2e^{2f}e^{-2p+2\mu} +e^{-f}e^{p-\mu} \right) \label{eq:TzitD3} \\
\bar{D} =& \frac{1}{\sigma}\left( e^{-2f}e^{-2\mu} +4(e^q+e^{-q})e^{f}e^{\mu} \right) \label{eq:TzitDbar3}
\end{align}
as the solutions to eq.\eqref{eq:Tzitmcc}.

There is also some freedom to redefine the external fields. We can shift $u$ or $v$ by an integer multiple of $2\pi i$ without affecting the bulk Lagrangians or the kinetic part of the defect Lagrangian. Taking $u\rightarrow u+2\pi i n$, $v\rightarrow v+2\pi i m$ (so $p\rightarrow p+\pi i (n+m)$, $q\rightarrow q+\pi i (n-m)$) gives the defect potential
\begin{align}
D =& \sigma \left( \frac{1}{2} e^{2f} (e^{2q}+e^{-2q}+2) e^{-2p+2\mu} +(-1)^{n+m} e^{-f} e^{p-\mu} \right) \\
\bar{D} =& \frac{1}{\sigma} \left( e^{-2f} e^{-2\mu} +4 (-1)^{n-m} e^{f} (e^q+e^{-q}) e^{\mu} \right)
.\end{align}
But we can also immediately take the redefinition $\mu\rightarrow\mu+\pi i (n+m)$ to return to the $D$ and $\bar{D}$ given in eqs.\eqref{eq:TzitD3}, \eqref{eq:TzitDbar3}, and since the freedom to shift the external fields corresponds to a shift in the auxiliary fields the entire family of momentum conserving defects satisfying the momentum conservation condition in eq.\eqref{eq:Tzitmcc} have a potential given by eqs.\eqref{eq:TzitD3}, \eqref{eq:TzitDbar3}.

The interactions of solitons with this defect were investigated in \cite{cz09b}, and a similar situation to the $A_r$ ATFT case was found, with the defect able to delay or absorb solitons and change their topological charge.

\subsection{Momentum conserving defects in the $D_4$ ATFT} \label{sec:mcd d4}

Here we present a more complete description of a defect in a $D_4$ ATFT, expanding on work carried out in \cite{bb17}.

The $D_4$ ATFT potential is given by eq.\eqref{eq:Lbulk} with simple (and lowest weight) roots
\begin{align}
\alpha_0 =& \left(\begin{matrix}
-1 \\
-1 \\
0 \\
0
\end{matrix}\right)
&
\alpha_1 =& \left(\begin{matrix}
1 \\
-1 \\
0 \\
0
\end{matrix}\right)
&
\alpha_2 =& \left(\begin{matrix}
0 \\
1 \\
-1 \\
0
\end{matrix}\right)
&
\alpha_{3} =& \left(\begin{matrix}
0 \\
0 \\
1 \\
-1
\end{matrix}\right)
&
\alpha_{4} =& \left(\begin{matrix}
0 \\
0 \\
1 \\
1
\end{matrix}\right) \label{eq:D4simpleroots}
\end{align}
and marks
\begin{align}
n_0 =& 1 & n_1 =& 1 & n_2 =& 2 & n_3 =& 1 & n_4 =& 1 \label{eq:D4marks}
.\end{align}
The fundamental weights $w_j$ satisfy $\langle\alpha_i,w_j\rangle=\delta_{ij}$, with $w_i$ being the fundamental weight associated to the simple root $\alpha_i$ and the fundamental weights of $D_4$ are
\begin{align}
w_1 =& \left(\begin{matrix}
1 \\
0 \\
0 \\
0
\end{matrix}\right)
&
w_2 =& \left(\begin{matrix}
1 \\
1 \\
0 \\
0
\end{matrix}\right)
&
w_3 =& \frac{1}{2} \left(\begin{matrix}
1 \\
1 \\
1 \\
-1
\end{matrix}\right)
&
w_4 =& \frac{1}{2} \left(\begin{matrix}
1 \\
1 \\
1 \\
1
\end{matrix}\right) \label{eq:D4fundamentalweights}
.\end{align}

In \cite{bb17} it was found that taking the 1-space to have the basis $(e_1,e_4)$ and the 2-space to have the basis $(e_2,e_3)$, giving two auxiliary fields $\mu_2$ and $\mu_3$, and taking $A=0$ and no $\xi$ fields gave a defect which, with the correct choice of potential, was momentum conserving. With these choices of 1-space and 2-space the defect Lagrangian in eq.\eqref{eq:finaldefectL} becomes
\begin{align}
\mathcal{L}^D =& u_1v_{1,t}+u_2v_{2,t}+u_3v_{3,t}+u_4v_{4,t}+2\mu_2\left(u_{2,t}-v_{2,t}\right)+2\mu_3\left(u_{3,t}-v_{3,t}\right) -D-\bar{D} \label{eq:LD4}
\end{align}
where $D(p_1,p_2-\mu_2,p_3-\mu_3,p_4,q_2,q_3)$ and $\bar{D}(q_1,q_2,q_3,q_4,\mu_2,\mu_3)$ (with $p_i=\frac{1}{2}(u_i+v_i)$, $q_i=\frac{1}{2}(u_i-v_i)$) must satisfy
\begin{align}
2\big( &
e^{-p_1-q_1-p_2-q_2}-e^{-p_1+q_1-p_2+q_2}
+e^{p_1+q_1-p_2-q_2}-e^{p_1-q_1-p_2+q_2}
+e^{p_2+q_2-p_3-q_3}-e^{p_2-q_2-p_3+q_3} \nonumber \\
&+e^{p_3+q_3-p_4-q_4}-e^{p_3-q_3-p_4+q_4}
+e^{p_3+q_3+p_4+q_4}-e^{p_3-q_3+p_4-q_4}\big) \nonumber \\
=& D_{p_1}\bar{D}_{q_1}+D_{q_2}\bar{D}_{\mu_2}-D_{\mu_2}\bar{D}_{q_2}+D_{q_3}\bar{D}_{\mu_3}-D_{\mu_3}\bar{D}_{q_3}+D_{p_4}\bar{D}_{q_4} \label{eq:D4mcc}
.\end{align}
From eqs.\eqref{eq:Dgeneral}, \eqref{eq:Dbargeneral} we expect $D$ and $\bar{D}$ to be
\begin{align}
D = \sigma\big( &
x_{0}(q_2,q_3)e^{-p_1-p_2+\mu_2}
+x_{1}(q_2,q_3)e^{p_1-p_2+\mu_2}
+x_{2}(q_2,q_3)e^{p_2-p_3-\mu_2+\mu_3} \nonumber \\
&+x_{3}(q_2,q_3)e^{p_3-p_4-\mu_3}
+x_{4}(q_2,q_3)e^{p_3+p_4-\mu_3}
\big) \label{eq:D4Dgeneralx} \\
\bar{D} = \frac{1}{\sigma}\big( &
y_{0}(q_1,q_2,q_3,q_4)e^{-\mu_2}
+y_{1}(q_1,q_2,q_3,q_4)e^{-\mu_2}
+y_{2}(q_1,q_2,q_3,q_4)e^{\mu_2-\mu_3} \nonumber \\
&+y_{3}(q_1,q_2,q_3,q_4)e^{\mu_3}
+y_{4}(q_1,q_2,q_3,q_4)e^{\mu_3}
\big) \label{eq:D4Dbargeneraly}
\end{align}
where $x_i$ and $y_i$ are unknown functions. As some terms in $\bar{D}$ have the same exponentials of $\mu$ we can redefine some of these currently arbitrary functions as $y_1\rightarrow y_1-y_0$ and $y_3\rightarrow y_3-y_4$ to set $y_0=0$ and $y_4=0$. We can also use the field redefinitions $\mu_2\rightarrow\mu_2-\left(\int^{q_2}\ln{x_0(q_2',q_3)}\mathrm{d}q_2'\right)_{q_2}$ and $\mu_3\rightarrow\mu_3-\left(\int^{q_2}\ln{x_0(q_2',q_3)}\mathrm{d}q_2'\right)_{q_3}$ to set $x_0=1$. The rest of the $x_i$ and $y_i$ can simply be redefined to include this extra function.

Using these choices in eq.\eqref{eq:xymcc} and equating powers of $\mu_{2,3}$ we find a set of differential equations which $x_i$ and $y_i$ must satisfy as a momentum conservation condition. While a single possible defect potential was given for the $D_4$ ATFT in \cite{bb17}, these differential equations were not solved exhaustively there, and the following working is new.

There are two distinct solutions,
\begin{align}
x_0 =& 1 & & \nonumber \\
x_1 =& 1 & y_1 =& \left(e^{q_1}+e^{-q_1}\right)\left(e^{q_2}+e^{-q_2}\right) \nonumber \\
x_2 =& 2g(q_3)\left(e^{q_2}+e^{-q_2}\right) & y_2 =& g(q_3)^{-1}\left(e^{q_3}+e^{-q_3}\right) \nonumber \\
x_3 =& \frac{1}{c} g(q_3)^{-1}\left(e^{q_3}+e^{-q_3}\right) & y_3 =& cg(q_3)\left(e^{q_4}+e^{-q_4}\right) \nonumber \\
x_4 =& \frac{1}{c} g(q_3)^{-1}\left(e^{q_3}+e^{-q_3}\right) & & \label{eq:D4xy1}
\end{align}
and
\begin{align}
x_0 =& 1 & & \nonumber \\
x_1 =& -1 & y_1 =& \left(e^{q_1}-e^{-q_1}\right)\left(e^{q_2}-e^{-q_2}\right) \nonumber \\
x_2 =& -2g(q_3)\left(e^{q_2}-e^{-q_2}\right) & y_2 =& g(q_3)^{-1}\left(e^{q_3}-e^{-q_3}\right) \nonumber \\
x_3 =& -\frac{1}{c} g(q_3)^{-1}\left(e^{q_3}-e^{-q_3}\right) & y_3 =& cg(q_3)\left(e^{q_4}-e^{-q_4}\right) \nonumber \\
x_4 =& \frac{1}{c} g(q_3)^{-1}\left(e^{q_3}-e^{-q_3}\right) & & \label{eq:D4xy2}
\end{align}
where the constant $c$ and function $g(q_3)$ are free (and may be different in each case). When used to write down $D$ and $\bar{D}$ from eqs.\eqref{eq:D4Dgeneralx}, \eqref{eq:D4Dbargeneraly} these will give two separate possibilities for the momentum conserving defect potential.

We can use our freedom to carry out field redefinitions to remove the constant $c$ and function $g$ in both cases. For the first solution taking $\mu_2\rightarrow\mu_2-\frac{1}{3}\ln{c}$, $\mu_3\rightarrow\mu_3-\frac{2}{3}\ln{c}$ and $\sigma\rightarrow c^{\frac{1}{3}}\sigma$ removes (or absorbs into the definition of $\mu^{(2)}$ and $\sigma$) the constant $c$ and taking $\mu_2\rightarrow\mu_2$, $\mu_3\rightarrow\mu_3-\ln{g(q_3)}$ removes the function $g(q_3)$. Reintroducing all possible freedom available from auxiliary field redefinitions by taking $\mu_2\rightarrow \mu_2+f(q_2,q_3)_{q_2}$, $\mu_3\rightarrow \mu_3+f(q_2,q_3)_{q_3}$ (where $f$ may be any function) we now have, from the first set of solutions, the defect potential
\begin{align}
D^+ =& \sigma\bigg(
e^{f_{q_2}}\left(e^{p_1}+e^{-p_1}\right)e^{-p_2+\mu_2}
+2e^{-f_{q_2}+f_{q_3}}\left(e^{q_2}+e^{-q_2}\right)e^{p_2-p_3-\mu_2+\mu_3} \nonumber \\
&\quad +e^{-f_{q_3}}\left(e^{q_3}+e^{-q_3}\right)\left(e^{p_4}+e^{-p_4}\right)e^{p_3-\mu_3} \label{eq:D4Dgeneral1}
\bigg) \\
\bar{D}^+ =& \frac{1}{\sigma}\bigg(
e^{-f_{q_2}}\left(e^{q_1}+e^{-q_1}\right)\left(e^{q_2}+e^{-q_2}\right)e^{-\mu_2}
+e^{f_{q_2}-f_{q_3}}\left(e^{q_3}+e^{-q_3}\right)e^{\mu_2-\mu_3} \nonumber \\
&\quad +e^{f_{q_3}}\left(e^{q_4}+e^{-q_4}\right)e^{\mu_3}
\bigg) \label{eq:D4Dbargeneral1}
.\end{align}
The $+$ superscripts will differentiate this from the defect potential arising from the second set of solutions, and refer to the fact that terms of the form $(e^q+e^{-q})$ appear here.

For the second solution taking $\mu_2\rightarrow\mu_2-\frac{1}{3}\ln{c}$, $\mu_3\rightarrow\mu_3-\frac{2}{3}\ln{c}$, $\sigma\rightarrow c^{\frac{1}{3}}\sigma$ and $\mu_3\rightarrow\mu_3-\ln{g(q_3)}$ again removes the constant $c$ and function $g(q_3)$. Reintroducing all possible freedom available from auxiliary field redefinitions by taking $\mu_2\rightarrow \mu_2+f(q_2,q_3)_{q_2}$, $\mu_3\rightarrow \mu_3+f(q_2,q_3)_{q_3}$ (where $f$ may be any function) we now have, from the second set of solutions, the defect potential
\begin{align}
D^- =& \sigma\bigg(
e^{f_{q_2}}\left(e^{p_1}-e^{-p_1}\right)e^{-p_2+\mu_2}
-2e^{-f_{q_2}+f_{q_3}}\left(e^{q_2}-e^{-q_2}\right)e^{p_2-p_3-\mu_2+\mu_3} \nonumber \\
&\quad +e^{-f_{q_3}}\left(e^{q_3}-e^{-q_3}\right)\left(e^{p_4}-e^{-p_4}\right)e^{p_3-\mu_3}
\bigg) \label{eq:D4Dgeneral2} \\
\bar{D}^- =& \frac{1}{\sigma}\bigg(
-e^{-f_{q_2}}\left(e^{q_1}-e^{-q_1}\right)\left(e^{q_2}-e^{-q_2}\right)e^{-\mu_2}
+e^{f_{q_2}-f_{q_3}}\left(e^{q_3}-e^{-q_3}\right)e^{\mu_2-\mu_3} \nonumber \\
&\quad +e^{f_{q_3}}\left(e^{q_4}-e^{-q_4}\right)e^{\mu_3}
\bigg) \label{eq:D4Dbargeneral2}
.\end{align}
The $-$ superscripts here refer to the fact that terms of the form $(e^q-e^{-q})$ appear.

There is still the freedom to carry out field redefinitions on the bulk fields. The bulk fields may be shifted by any $2\pi i$ multiple of a weight of $D_4$ without affecting the bulk Lagrangians. If $u$ and $v$ have the same shift then $p$ is also shifted by a $2\pi i$ multiple of a weight, and as exponentials of $p$ in $D$ all appear in the form $e^{\alpha_i.p}$ they remain unchanged. $q$ remains completely unchanged. So as in the Tzitz\'{e}ica case it is the relative shift between $u$ and $v$ which is important. We will consider shifts of $v$ proportional to the fundamental weights given in eqs.\eqref{eq:D4fundamentalweights}.

Acting on the defect potential given by $D^+$, $\bar{D}^+$ in eqs.\eqref{eq:D4Dgeneral1}, {eq:D4Dbargeneral1} with $v\rightarrow v+2\pi i w_1$, where $w_1$ is one of the fundamental weights given in eq.\eqref{eq:D4fundamentalweights}, and also performing the shift $\mu_3\rightarrow\mu_3+\pi i$ on the auxiliary fields and the redefinition $\sigma\rightarrow-\sigma$ gives $D^+$, $\bar{D}^+$. The freedom from this external field redefinition is equivalent to the freedom we already have to redefine the auxiliary fields and the defect parameter, and does not give a defect potential that is materially different. Carrying out an identical set of redefinitions on $D^-$, $\bar{D}^-$ returns to $D^-$, $\bar{D}^-$ also.

Acting on $D^+$, $\bar{D}^+$ with $v\rightarrow v+2\pi i w_2$ immediately returns $D^+$, $\bar{D}^+$, and likewise acting on $D^-$, $\bar{D}^-$ with $v\rightarrow v+2\pi i w_2$ immediately returns $D^-$, $\bar{D}^-$.

Acting on $D^+$, $\bar{D}^+$ with $v\rightarrow v+2\pi i w_3$ and $\mu_3\rightarrow\mu_3-\frac{\pi i}{2}$ gives $D^-$, $\bar{D}^-$, so the two defect potentials, while not linked by any redefinitions of the auxiliary fields, are linked by a shift of the bulk fields. Using the same shift and set of redefinitions on $D^-$, $\bar{D}^-$ returns $D^+$, $\bar{D}^+$.

Finally acting on $D^+$, $\bar{D}^+$ with $v\rightarrow v+2\pi i w_4$, the shifts $\mu_2\rightarrow\mu_2+\pi i$, $\mu_3\rightarrow\mu_3-\frac{\pi i}{2}$ and the redefinition $\sigma\rightarrow-\sigma$ gives $D^-$, $\bar{D}^-$. Unsurprisingly the same set of field redefinitions take $D^-$, $\bar{D}^-$ to $D^+$, $\bar{D}^+$.

A shift of a $2\pi i$ multiple of fundamental weights $w_{1,2}$ has no effect on either defect potential beyond utilising the freedom to make auxiliary field redefinitions which is already encapsulated by the presence of the arbitrary function $f$ in the potentials. A shift which is a $2\pi i$ multiple of fundamental weights $w_{3,4}$ links the two distinct defect potentials.

\section{Zero curvature for systems with defects}

We have now given all the necessary background on the generalised type II defects from \cite{bb17}. In this section we first give the defect zero curvature condition, then apply it to the defects given in sections \ref{sec:mcd}-\ref{sec:mcd d4}.

\subsection{General defect zero curvature condition} \label{sec:zcc}

Consider a defect at $x=0$. There will be an integrable theory in the region $x\leq 0$ with the Lax pair $a_0^<(t,x)$, $a_1^<(t,x)$ dependent on the field $u$ and satisfying the zero curvature condition in eq.\eqref{eq:zcc}, and an integrable theory in the region $x\geq 0$ with the Lax pair $a_0^>(t,x)$, $a_1^>(t,x)$ dependent on the field $v$ and also satisfying eq.\eqref{eq:zcc}. We consider the transport of the vector $\Psi$ in the region of the defect, where some time dependent defect matrix $K$ acts to move from the left of the defect to the right of the defect without changing position.
\begin{align}
\includegraphics{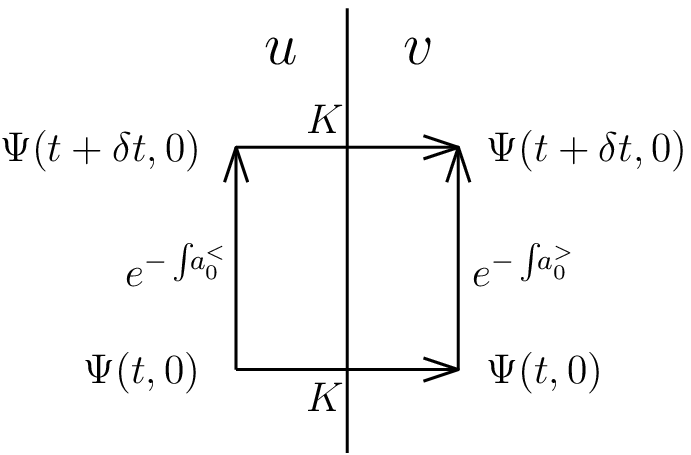}
\end{align}
The defect transport matrix $K$ depends on both the $u$ and $v$ fields evaluated at $x=0$ and on any auxiliary fields which are confined to the defect. The Lax matrices on either side of the defect will be dependent on the same spectral parameter $\lambda$, and $K$ will also be dependent on $\lambda$. These Lax matrices and the defect transport matrix $K$ can then be used together to give the monodromy matrix which transports $\Psi$ from $x\rightarrow-\infty$ to $x\rightarrow\infty$. For this transport to be path independent we require
\begin{align}
K(t+\delta t)
Pe^{-\int_t^{t+\delta t} \mathrm{d}t' a_0^<(t',0)}
=&
Pe^{-\int_t^{t+\delta t} \mathrm{d}t' a_0^>(t',0)}
K(t)
\end{align}
and expanding this in $\delta t$ we have
\begin{align}
K_t=Ka_0^<-a_0^>K \label{eq:defectzcc}
\end{align}
evaluated at $x=0$. This calculation of the defect zero curvature condition is not specific to defects in ATFTs, but can be applied to a defect in any integrable theory. The zero curvature condition is the same as that found in \cite{bcz04b}.

The bulk zero curvature condition in eq.\eqref{eq:zcc} is satisfied if and only if the bulk equations of motion are satisfied, and this extra defect zero curvature condition must be satisfied if and only if the defect equations are satisfied. Note that eq.\eqref{eq:defectzcc} is equivalent to $K$ being a gauge transformation between the operators $\partial_t+a_0^<$ and $\partial_t+a_0^>$, with $\partial_t+a_0^<=K^{-1}(\partial_t+a_0^>)K$. Carrying out a gauge transform of $G$ on $a_0^<$ and $G'$ on $a_0^>$ (as given in eq.\eqref{eq:a0GT}) along with the gauge transformation $K\rightarrow K'=G'KG^{-1}$ leaves this defect zero curvature condition unchanged.

\subsection{Zero curvature for a defect in an ATFT} \label{sec:zcc atft}

Using the ATFT $a_0$ matrix given in eq.\eqref{eq:ATFTa0} and taking it to depend on $u=p+q$ to give $a_0^<$ and $v=p-q$ to give $a_0^>$ the zero curvature condition on the defect becomes
\begin{align}
2K_t =&
p_{j,x} \left[ K,H_j \right]
+q_{j,x} \{ K,H_j \} \nonumber \\
&+\frac{1}{\sqrt{2}} \sum_{i=0}^r \sqrt{n_i}|\alpha_i| e^{\frac{1}{2}(\alpha_i)_jp_j}
\bigg(
\lambda\left(
e^{\frac{1}{2}(\alpha_i)_jq_j} KE_{\alpha_i} -e^{-\frac{1}{2}(\alpha_i)_jq_j} E_{\alpha_i}K
\right) \nonumber \\
&\hspace{11.2em} -\frac{1}{\lambda}\left(
e^{\frac{1}{2}(\alpha_i)_jq_j} KE_{-\alpha_i} -e^{-\frac{1}{2}(\alpha_i)_jq_j} E_{-\alpha_i}K
\right)\!
\bigg) \label{eq:Kgeneral1}
\end{align}
where square brackets indicate a commutator and curly brackets an anticommutator (not a Poisson bracket).

We will begin by taking the defect to be of the general form given in eq.\eqref{eq:finaldefectL}, which has defect equations given in eqs.\eqref{eq:defecteom1}-\eqref{eq:defecteom6},
where $D$ and $\bar{D}$ must have the dependencies given in eqs.\eqref{eq:Ddependencies}, \eqref{eq:Dbardependencies} and satisfy the additional momentum conservation condition in eq.\eqref{eq:mcc}. Using eqs.\eqref{eq:defecteom1}-\eqref{eq:defecteom4} to remove all $x$ derivatives from eq.\eqref{eq:Kgeneral1} gives
\begin{align}
2K_t =&
\left(p^{(1)}_{j,t} -2q^{(1)}_{k,t}A_{kj} -\frac{1}{2}D_{q^{(1)}_j} -\frac{1}{2}\bar{D}_{q^{(1)}_j}\right) \left[ K,H^{(1)}_j \right] \nonumber \\
&+\left(p^{(2)}_{j,t} -2\mu^{(2)}_{j,t} -\frac{1}{2}D_{q^{(2)}_j} -\frac{1}{2}\bar{D}_{q^{(2)}_j}\right) \left[ K,H^{(2)}_j \right] \nonumber \\
&+\left(-q^{(1)}_{j,t} -\frac{1}{2}D_{p^{(1)}_j}\right) \{ K,H^{(1)}_j \}
+\left(-q^{(2)}_{j,t} -\frac{1}{2}D_{p^{(2)}_j}\right) \{ K,H^{(2)}_j \} \nonumber \\
&+\frac{1}{\sqrt{2}} \sum_{i=0}^r \sqrt{n_i}|\alpha_i| e^{\frac{1}{2}(\alpha_i)_jp_j}
\bigg(
\lambda\left(
e^{\frac{1}{2}(\alpha_i)_jq_j} KE_{\alpha_i} -e^{-\frac{1}{2}(\alpha_i)_jq_j} E_{\alpha_i}K
\right) \nonumber \\
&\hspace{11.1em} -\frac{1}{\lambda}\left(
e^{\frac{1}{2}(\alpha_i)_jq_j} KE_{-\alpha_i} -e^{-\frac{1}{2}(\alpha_i)_jq_j} E_{-\alpha_i}K
\right) \!
\bigg) \label{eq:Kgeneral2}
.\end{align}
Every Cartan generator is associated with one of the orthonormal basis vectors of the root space, so $H^{(1)}$ denotes the Cartan generators which are associated with the orthonormal basis vectors which form a basis of the 1-space and $H^{(2)}$ denotes the Cartan generators associated with the orthonormal basis vectors of the 2-space. The $t$ derivatives on the right hand side can be removed by applying the transformation
\begin{align}
K=e^{-\frac{1}{2}(p_{j}+q_{j})H_{j}+q^{(1)}{j}A_{jk}H^{(1)}_{k} +\mu^{(2)}_jH^{(2)}_{j}}\hat{K}e^{\frac{1}{2}(p_{j}-q_{j})H_{j}-q^{(1)}_{j}A_{jk}H^{(1)}_{k}-\mu^{(2)}_jH^{(2)}_{j}}
\end{align}
to give
\begin{align}
&4\hat{K}_{t}
+D_{p_{j}} \{\hat{K},H_j\}
+(D_{q_{j}}+\bar{D}_{q_{j}}) \left[\hat{K},H_j\right] \nonumber \\
&=
\sqrt{2}\sum_{i=0}^r \sqrt{n_i} |\alpha_i|
\bigg( \lambda e^{(\alpha_i)_{j}p_{j} +(\alpha_i)^{(1)}_{j}A_{jk}q^{(1)}_{k}-(\alpha_i)^{(2)}_{j}\mu^{(2)}_j}
\left[\hat{K},E_{\alpha_i}\right] \nonumber \\
&\qquad\qquad
-\frac{1}{\lambda} e^{-(\alpha_i)^{(1)}_{j}A_{jk}q^{(1)}_{k}+(\alpha_i)^{(2)}_{j}\mu^{(2)}_j} \left( e^{(\alpha_i)_{j}q_{j}}\hat{K}E_{-\alpha_i} -e^{-(\alpha_i)_{j}q_{j}} E_{-\alpha_i}\hat{K} \right) \bigg) \label{eq:Kgeneral3}
.\end{align}

If $\hat{K}$ is dependent on a field then the term $\hat{K}_t$ introduces a $t$ derivative of that field, which will not appear anywhere else in eq.\eqref{eq:Kgeneral3}. For the fields $q^{(2)}$ and $\xi$ we can remove the $t$ derivative using eq.\eqref{eq:defecteom5} and eq.\eqref{eq:defecteom6} respectively. For the fields $p^{(1)}$, $q^{(1)}$, $p^{(2)}$ and $\mu^{(2)}$ the $t$ derivative cannot be removed (except by the introduction of an $x$ derivative, which returns us to the previous step in our calculation) so $\hat{K}$ cannot be dependent on these fields. The same argument can be used to show that $\hat{K}$ cannot depend on the derivatives of fields as well. With $\hat{K}$ only dependent on $q^{(2)}$ and $\xi$ we have $\hat{K}_t=\hat{K}_{q^{(2)}_i}q^{(2)}_{i,t}+\hat{K}_{\xi_i}\xi_{i,t}$, and using this and eqs.\eqref{eq:defecteom5}, \eqref{eq:defecteom6} the zero curvature condition becomes
\begin{align}
&\hat{K}_{q^{(2)}_i}(D_{\mu^{(2)}_{i}}+\bar{D}_{\mu^{(2)}_{i}})
-4\hat{K}_{\xi_i}W_{ij}(D_{\xi_j}+\bar{D}_{\xi_j})
+D_{p_{j}} \{\hat{K},H_j\}
+(D_{q_{j}}+\bar{D}_{q_{j}}) \left[\hat{K},H_j\right] \nonumber \\
&=
\sqrt{2}\sum_{i=0}^r \sqrt{n_i} |\alpha_i|
\bigg( \lambda e^{(\alpha_i)_{j}p_{j} +(\alpha_i)^{(1)}_{j}A_{jk}q^{(1)}_{k}-(\alpha_i)^{(2)}_{j}\mu^{(2)}_j}
\left[\hat{K},E_{\alpha_i}\right] \nonumber \\
&\qquad\qquad
-\frac{1}{\lambda} e^{-(\alpha_i)^{(1)}_{j}A_{jk}q^{(1)}_{k}+(\alpha_i)^{(2)}_{j}\mu^{(2)}_j} \left( e^{(\alpha_i)_{j}q_{j}}\hat{K}E_{-\alpha_i} -e^{-(\alpha_i)_{j}q_{j}} E_{-\alpha_i}\hat{K} \right) \bigg) \label{eq:Kgeneral4}
.\end{align}

To progress further we now need a specific form for the defect potential. In section \ref{sec:mcd atft} we stated that for a defect in an ATFT to be momentum conserving $D$ and $\bar{D}$ must be of the form given in eqs.\eqref{eq:Dgeneral}, \eqref{eq:Dbargeneral}. Using this in the zero curvature condition we have
\begin{align}
&\sigma \sum_{i=0}^r
e^{(\alpha_i)_jp_j +(\alpha_i)^{(1)}_jA_{jk}q^{(1)}_k -(\alpha_i)^{(2)}_j\mu^{(2)}_j)}
\big(
-x_i (\alpha_i)^{(2)}_j \hat{K}_{q^{(2)}_j}
+4x_{i,\xi_j} W_{jk} \hat{K}_{\xi_k} \nonumber \\
&\qquad\quad +x_i (\alpha_i)^{(1)}_j A_{jk} \left[\hat{K},H^{(1)}_k\right]
+x_{i,q^{(2)}_j} \left[\hat{K},H^{(2)}_j\right]
+x_i (\alpha_i)_j \{\hat{K},H_j\}
\big) \nonumber \\
&+\frac{1}{\sigma} \sum_{i=0}^r
e^{-(\alpha_i)^{(1)}_jA_{jk}q^{(1)}_k +(\alpha_i)^{(2)}_j\mu^{(2)}_j}
\big(
y_i (\alpha_i)^{(2)}_j \hat{K}_{q^{(2)}_j}
+4y_{i,\xi_j} W_{jk} \hat{K}_{\xi_k} \nonumber \\
&\qquad\qquad -y_i (\alpha_i)^{(1)}_j A_{jk} \left[\hat{K},H^{(1)}_k\right]
+y_{i,q_j} \left[\hat{K},H_j\right]
\big) \nonumber \\
& =
\sqrt{2}\sum_{i=0}^r \sqrt{n_i} |\alpha_i|
\bigg( \lambda e^{(\alpha_i)_{j}p_{j} +(\alpha_i)^{(1)})_{j}A_{jk}q^{(1)}_{k}-(\alpha_i)^{(2)}_{j}\mu^{(2)}_j}
\left[\hat{K},E_{\alpha_i}\right] \nonumber \\
&\qquad\qquad
-\frac{1}{\lambda} e^{-(\alpha_i)^{(1)}_{j}A_{jk}q^{(1)}_{k}+(\alpha_i)^{(2)}_{j}\mu^{(2)}_j} \left( e^{(\alpha_i)_{j}q_{j}}\hat{K}E_{-\alpha_i} -e^{-(\alpha_i)_{j}q_{j}} E_{-\alpha_i}\hat{K} \right) \bigg)
.\end{align}
Equating exponents of $p$ splits this into $r+2$ equations,
\begin{align}
\sqrt{2} \sqrt{n_i} |\alpha_i|
\rho \left[\hat{K},E_{\alpha_i}\right] =&
-x_i (\alpha_i)^{(2)}_j \hat{K}_{q^{(2)}_j}
+4x_{i,\xi_j} W_{jk} \hat{K}_{\xi_k} \nonumber \\
&+x_i (\alpha_i)^{(1)}_j A_{jk} \left[\hat{K},H^{(1)}_k\right]
+x_{i,q^{(2)}_j} \left[\hat{K},H^{(2)}_j\right] \nonumber \\
&+x_i (\alpha_i)_j \{\hat{K},H_j\} \label{eq:Khat1}
\end{align}
for $i=0,\dots,r$ and
\begin{align}
&-\sqrt{2}\sum_{i=0}^r \sqrt{n_i} |\alpha_i|
e^{-(\alpha_i)^{(1)}_{j}A_{jk}q^{(1)}_{k}+(\alpha_i)^{(2)}_{j}\mu^{(2)}_j} \left( e^{(\alpha_i)_{j}q_{j}}\hat{K}E_{-\alpha_i} -e^{-(\alpha_i)_{j}q_{j}} E_{-\alpha_i}\hat{K} \right) \nonumber \\
&\qquad=
\rho \sum_{i=0}^r
e^{-(\alpha_i)^{(1)}_jA_{jk}q^{(1)}_k +(\alpha_i)^{(2)}_j\mu^{(2)}_j}
\Big(
-y_i (\alpha_i)^{(1)}_j A_{jk} \left[\hat{K},H^{(1)}_k\right]
+y_{i,q_j} \left[\hat{K},H_j\right] \nonumber \\
&\hspace{16em} +y_i (\alpha_i)^{(2)}_j \hat{K}_{q^{(2)}_j}
+4y_{i,\xi_j} W_{jk} \hat{K}_{\xi_k}
\Big) \label{eq:Khat2}
\end{align}
where we have set $\rho=\lambda\sigma^{-1}$. We cannot split eq.\eqref{eq:Khat2} by equating exponentials of $\mu^{(2)}$, as two different roots $\alpha_i$ amd $\alpha_j$ may have the same projection onto the 2-space.

Multiplying $K$ by a constant does not affect the zero curature condition in eq.\eqref{eq:defectzcc}, so we can always take the highest power of $\rho$ appearing in $K$ to be zero. Therefore we can always expand $\hat{K}$ in $\rho$ as
\begin{align}
\hat{K} =& \sum_{s=0}^{\infty} \rho^{-s}k_s \label{eq:Kexpansion}
.\end{align}
The $k_s$ are matrices, and any of them may be zero. We do not know if this expansion terminates. We will assume that, like the bulk Lax pair, this defect matrix will consist of generators of the Lie algebra. More specifically, since it appears as part of the monodromy matrix, we would expect to be able to write it as an exponential or combination of exponentials of the generators. Expanding such an exponential in terms of $\rho$ (which should appear in the exponent by comparison with the bulk monodromy matrix) we therefore expect that the matrices $k_s$ will be some combination of generator matrices.

Substituting this expansion into the zero curvature relations in eqs.\eqref{eq:Khat1}, \eqref{eq:Khat2} and equating powers of $\rho$ gives a set of recursion relations,
\begin{align}
\sqrt{2} \sqrt{n_i} |\alpha_i|
\left[k_{s+1},E_{\alpha_i}\right] =&
-x_i (\alpha_i)^{(2)}_j k_{s,q^{(2)}_j}
+4x_{i,\xi_j} W_{jk} k_{s,\xi_k} \nonumber \\
&+x_i (\alpha_i)^{(1)}_j A_{jk} \left[k_s,H^{(1)}_k\right]
+x_{i,q^{(2)}_j} \left[k_s,H^{(2)}_j\right]
+x_i (\alpha_i)_j \{k_s,H_j\} \label{eq:krecursion1}
\end{align}
for $i=0,\dots,r$ and
\begin{align}
&-\sqrt{2}\sum_{i=0}^r \sqrt{n_i} |\alpha_i|
e^{-(\alpha_i)^{(1)}_{j}A_{jk}q^{(1)}_{k}+(\alpha_i)^{(2)}_{j}\mu^{(2)}_j} \left( e^{(\alpha_i)_{j}q_{j}}k_sE_{-\alpha_i} -e^{-(\alpha_i)_{j}q_{j}} E_{-\alpha_i}k_s \right) \nonumber \\
&\qquad=
\sum_{i=0}^r
e^{-(\alpha_i)^{(1)}_jA_{jk}q^{(1)}_k +(\alpha_i)^{(2)}_j\mu^{(2)}_j}
\Big(-y_i (\alpha_i)^{(1)}_j A_{jk} \left[k_{s+1},H^{(1)}_k\right]
+y_{i,q_j} \left[k_{s+1},H_j\right]
\nonumber \\
&\qquad\qquad\qquad\qquad\qquad\qquad\qquad \quad
+y_i (\alpha_i)^{(2)}_j k_{s+1,q^{(2)}_j}
+4y_{i,\xi_j} W_{jk} k_{s+1,\xi_k}
\Big) \label{eq:krecursion2}
.\end{align}

We can now attempt to solve these relations, which would ensure zero curvature across any momentum conserving defect of the form given in eq.\eqref{eq:finaldefectL} in an ATFT. Unfortunately it is not possible to solve the recursion relations for all values of $s$ for a general defect in an ATFT, but the $s=-1$, $s=0$ and $s=1$ recursion relations give us an idea of the form all $k_s$ matrices will take, and if the expansion terminates then the recursion relation for the highest value of $s$ gives some potentially useful constraints on the splitting of the root space into the 1-space and 2-space.

Beginning with $s=-1$ we have
\begin{align}
0 =&
\sqrt{2} \sqrt{n_i} |\alpha_i|
\left[k_{0},E_{\alpha_i}\right]  \label{eq:ks1}
\end{align}
for $i= 0,\dots,r$ and
\begin{align}
0 =&
\sum_{i=0}^r
e^{-(\alpha_i)^{(1)}_jA_{jk}q^{(1)}_k +(\alpha_i)^{(2)}_j\mu^{(2)}_j}
\big( -y_i (\alpha_i)^{(1)}_j A_{jk} \left[k_{0},H^{(1)}_k\right]
+y_{i,q_j} \left[k_{0},H_j\right]
\nonumber \\
&\qquad\qquad\qquad\qquad\qquad\qquad +y_i (\alpha_i)^{(2)}_j k_{0,q^{(2)}_j}
+4y_{i,\xi_j} W_{jk} k_{0,\xi_k}
\big) \label{eq:ks2}
.\end{align}
If $k_0$ is to commute with all simple root generators and the lowest weight root generator then by Schur's lemma it must be proportional to the identity matrix. This ensures the first $r+1$ equations are satisfied. We will take $k_0$ to be a scalar multiple of the identity matrix (satisfying the final equation), and using the fact that $K$ may be multiplied by a constant without affecting the defect zero curvature condition, set $k_0=\mathbb{1}$. There may be some choices of $k_0$ which are dependent on $q^{(2)}$ and $\xi$ and satisfy eq.\eqref{eq:ks2}, but it is certainly not obvious. No defects found thus far have contained auxiliary fields which couple only to other auxiliary fields, and if these is no $\xi$ field vector then for eq.\eqref{eq:ks2} to be satisfied we must have $k_{0,q^{(2)}_i}=0$ and so $k_0$ will always be a scalar multiple of the identity matrix.

Now consider $s=0$. The recurrence relations give
\begin{align}
\sqrt{2} \sqrt{n_i} |\alpha_i|
\left[k_{1},E_{\alpha_i}\right] =&
2x_i (\alpha_i)_j H_j \label{eq:ks3}
\end{align}
for $i=0,\dots,r$ and
\begin{align}
&-\sqrt{2}\sum_{i=0}^r \sqrt{n_i} |\alpha_i|
e^{-(\alpha_i)^{(1)}_{j}A_{jk}q^{(1)}_{k}+(\alpha_i)^{(2)}_{j}\mu^{(2)}_j} \left( e^{(\alpha_i)_{j}q_{j}} -e^{-(\alpha_i)_{j}q_{j}} \right) E_{-\alpha_i} \nonumber \\
&\qquad=
\sum_{i=0}^r
e^{-(\alpha_i)^{(1)}_jA_{jk}q^{(1)}_k +(\alpha_i)^{(2)}_j\mu^{(2)}_j}
\big( -y_i (\alpha_i)^{(1)}_j A_{jk} \left[k_{1},H^{(1)}_k\right]
+y_{i,q_j} \left[k_{1},H_j\right]
\nonumber \\
&\qquad\qquad\qquad\qquad\qquad\qquad\qquad +y_i (\alpha_i)^{(2)}_j k_{1,q^{(2)}_j}
+4y_{i,\xi_j} W_{jk} k_{1,\xi_k}
\big) \label{eq:ks4}
,\end{align}
and we can immediately see that the first $r+1$ equations in eq.\eqref{eq:ks3} are satisfied by
\begin{align}
k_1 =& -\frac{1}{\sqrt{2}} \sum_{j=0}^r \frac{1}{\sqrt{n_j}} |\alpha_j| x_j E_{-\alpha_j}
\end{align}
using the fact that a simple root plus the negative of a simple root is never a root and that the highest (lowest) weight root plus any positive (negative) root cannot be a root. The final equation, eq.\eqref{eq:ks4}, then becomes
\begin{align}
&2\sum_{i=0}^r \sqrt{n_i} |\alpha_i|
e^{-(\alpha_i)^{(1)}_{k}A_{kl}q^{(1)}_{l}+(\alpha_i)^{(2)}_{k}\mu^{(2)}_k} \left( e^{(\alpha_i)_{k}q_{k}} -e^{-(\alpha_i)_{k}q_{k}} \right) E_{-\alpha_i} \nonumber \\
&=
\sum_{i=0}^r \sum_{j=0}^r
\frac{1}{\sqrt{n_j}} |\alpha_j|
e^{-(\alpha_i)^{(1)}_kA_{kl}q^{(1)}_l +(\alpha_i)^{(2)}_k\mu^{(2)}_k}
\big(
y_i (\alpha_i)^{(2)}_k x_{j,q^{(2)}_k}
+4y_{i,\xi_k} W_{kl} x_{j,\xi_l} +x_j y_{i,q_k} (\alpha_j)_k \nonumber \\
&\hspace{18em} -x_j y_i (\alpha_i)^{(1)}_k A_{kl} (\alpha_j)^{(1)}_l
\big)
E_{-\alpha_j}
\end{align}
where we have made use of eq.\eqref{eq:HEE}. Because the generators of the simple and lowest weight roots are linearly independent we can equate the coefficients of these matrices to give
\begin{align}
2 n_i
\left( e^{(\alpha_i)_{k}q_{k}} -e^{-(\alpha_i)_{k}q_{k}} \right)
=&\sum_{j=0}^r
e^{(\alpha_i-\alpha_j)^{(1)}_kA_{kl}q^{(1)}_l +(\alpha_j-\alpha_i)^{(2)}_k\mu^{(2)}_k} \nonumber \\
&\hspace{-9em}\big(
y_j (\alpha_j)^{(2)}_k x_{i,q^{(2)}_k}
+4y_{j,\xi_k} W_{kl} x_{i,\xi_l}
+x_i y_{j,q_k} (\alpha_i)_k
-x_i y_j (\alpha_j)^{(1)}_k A_{kl} (\alpha_i)^{(1)}_l
\big)
\end{align}
for $i=0,\dots,r$. But this is identical to the set of differential equations appearing in eq.\eqref{eq:xymcc}, which came from taking $D$ and $\bar{D}$ to be of the form in eqs.\eqref{eq:Dgeneral}, \eqref{eq:Dbargeneral} then substituting these into the momentum conservation condition in eq.\eqref{eq:mcc} to give a set of differential equations which must be satisfied by $x_i$ and $y_i$ if the defect is to be momentum conserving. We have not quite shown that momentum conservation is necessary for a system with a defect to have zero curvature, as we made the assumption that $k_0$ did not depend on $\xi$. We also have not shown that momentum conservation is a sufficient condition as this would require the recursion relations to be satisfied for all values of $s$. However, this highlights the link between momentum conservation and integrability, and for all defects found in \cite{bb17} their momentum conservation is necessary if they are to be integrable.

These first two terms indicate some sort of pattern of grading, with the $n^{th}$ power of $\rho$ in the expansion of $\hat{K}$ containing the product (or rather a sum of products) of $n$ generators $E_{-\alpha_i}$ ($i=0,\dots,r$). From eq.\eqref{eq:EEE} we see that the generators of roots which are not simple or the lowest weight root can still be written as a sum of products of the generators of simple or lowest weight roots. This also implies some cyclicity, as by taking commutators of $E_{-\alpha_0}$ with $E_{-\alpha_i}$ ($i=1,\dots,r$) we can eventually reach $H$. So the Cartan generators can be written as a sum of products of $\sum_{i=1}^r n_i+1$ generators of negatives of simple roots and the generator associated with the highest weight root. So (from eq.\eqref{eq:HEE}) the generators $E_{-\alpha_i}$ ($i=0,\dots,r$) can be written as a sum of products of $\sum_{i=1}^r n_i+2$ such generators. So if this grading pattern continues then the terms in the expansion in eq.\eqref{eq:Kexpansion} with $\rho^{-\sum n_i -1-i}$ are a rewriting of the terms with $\rho^{-i}$.

By inspection of the $s=1$ recursion relations it appears that the grading described here will give the correct matrices from the commutators appearing in the recursion relation. However, actually calculating $k_2$ is too difficult, as we do not know anything about the root structure of the underlying Lie algebra and so do not know the exact form of the commutation relations for the generators. To actually calculate this defect zero curvature matrix we will need to consider specific ATFTs.

However, there is still some useful information about defects in ATFTs to be gleaned from these recursion relations if we consider what happens if the expansion for $\hat{K}$ terminates. Let us assume that for all $s>n$ we have $k_s=0$. Then take $s=n$ for the recursion relations, giving
\begin{align}
0 =&
-x_i (\alpha_i)^{(2)}_j k_{n,q^{(2)}_j}
+4x_{i,\xi_j} W_{jk} k_{n,\xi_k} \nonumber \\
&+x_i (\alpha_i)^{(1)}_j A_{jk} \left[k_n,H^{(1)}_k\right]
+x_{i,q^{(2)}_j} \left[k_n,H^{(2)}_j\right]
+x_i (\alpha_i)_j \{k_n,H_j\} \label{eq:ks5}
\end{align}
for $i=0,\dots,r$ and
\begin{align}
&\sum_{i=0}^r \sqrt{n_i} |\alpha_i|
e^{-(\alpha_i)^{(1)}_{j}A_{jk}q^{(1)}_{k}+(\alpha_i)^{(2)}_{j}\mu^{(2)}_j} \left( e^{(\alpha_i)_{j}q_{j}}k_nE_{-\alpha_i} -e^{-(\alpha_i)_{j}q_{j}} E_{-\alpha_i}k_n \right) =0 \label{eq:ks6}
.\end{align}
We will not solve these equations, but can use eq.\eqref{eq:ks6} to get some information on the form of defects with zero curvature.

For the right hand side of eq.\eqref{eq:ks6} to be zero the terms appearing there must either be equal to zero or proportional to another term, enabling cancellations to occur. For a term to disappear $k_n$ must annihilate $E_{-\alpha_i}$ or vice versa. However, to know whether this happens and for which terms we need to know not just $k_n$ but also what the underlying Lie algebra is and what representation we are using. We will therefore assume that this is never the case, and so every term in eq.\eqref{eq:ks6} is non-zero. This assumption is acceptable as we are not trying to prove every defect with zero curvature must take a particular form. Instead we are looking for constraints which apply in certain cases which may be useful in finding momentum conserving defects for the $E$ series ATFTs, which were not covered by the trial-and-error method used in \cite{bb17}. 

Every term in eq.\eqref{eq:ks6} must cancel with at least one other term. First consider a cancellation between terms $k_nE_{-\alpha_i}$ and $k_nE_{-\alpha_j}$. Because $k_n$ is only dependent on $q^{(2)}$ and $\xi$ any dependence on $q^{(1)}$ and $\mu^{(2)}$ appearing in these two terms must match. From the exponentials appearing in these terms this requires
\begin{align}
(\alpha_i)^{(1)}_{k}q^{(1)}_{k}-(\alpha_i)^{(1)}_{k}A_{kl}q^{(1)}_{l}+(\alpha_i)^{(2)}_{k}\mu^{(2)}_k=(\alpha_j)^{(1)}_{k}q^{(1)}_{k}-(\alpha_j)^{(1)}_{k}A_{kl}q^{(1)}_{l}+(\alpha_j)^{(2)}_{k}\mu^{(2)}_k \label{eq:kncancel1}
.\end{align}
Because $A$ is real and antisymmetric the matrix $\mathbb{1}\pm A$ has complex eigenvalues which are all non-zero, so is invertible. Therefore requiring eq.\eqref{eq:kncancel1} to hold gives $\alpha_i=\alpha_j$, so we cannot have a cancellation between two terms of the form $k_nE_{-\alpha_i}$. Next consider a cancellation between terms $E_{-\alpha_i}k_n$ and $E_{-\alpha_j}k_n$. This requires
\begin{align}
-(\alpha_i)^{(1)}_{k}q^{(1)}_{k}-(\alpha_i)^{(1)}_{k}A_{kl}q^{(1)}_{l}+(\alpha_i)^{(2)}_{k}\mu^{(2)}_k=-(\alpha_j)^{(1)}_{k}q^{(1)}_{k}-(\alpha_j)^{(1)}_{k}A_{kl}q^{(1)}_{l}+(\alpha_j)^{(2)}_{k}\mu^{(2)}_k
,\end{align}
which again immediately gives $\alpha_i=\alpha_j$, and so no cancellations. So all cancellations must be between a term of the form $k_nE_{-\alpha_i}$ and another term of the form $E_{-\alpha_j}k_n$. This requires every root $\alpha_i$ to have another root $\alpha_j$ for which it satisfies
\begin{align}
(\alpha_i)^{(1)}_{k}q^{(1)}_{k}-(\alpha_i)^{(1)}_{k}A_{kl}q^{(1)}_{l}+(\alpha_i)^{(2)}_{k}\mu^{(2)}_k=-(\alpha_j)^{(1)}_{k}q^{(1)}_{k}-(\alpha_j)^{(1)}_{k}A_{kl}q^{(1)}_{l}+(\alpha_j)^{(2)}_{k}\mu^{(2)}_k
.\end{align}

If the assumptions we have made about the $\hat{K}$ series terminating and the $k_n$ matrix not annihilating any $E_{\alpha}$ operators hold (and for the Tzitz\'{e}ica and $D_4$ defect matrices we find in the following sections they do hold) then we have some fairly restrictive constraints on the projections of the roots onto the 1-space and 2-space. Either the root $\alpha_i$ must have $(\alpha_i)^{(1)}=0$, in which case the $k_nE_{\alpha_i}$ term is able to cancel with $E_{\alpha_i}k_n$, or there must be some other root $\alpha_j$ with $(\mathbb{1}+A)\alpha^{(1)}_i=(-\mathbb{1}+A)\alpha^{(1)}_j$ and $\alpha^{(2)}_i=\alpha^{(2)}_j$. By their projections onto the 2-space we should be able to find sets of roots whose projections onto the 1-space are linked.

For the $A_r$ ATFTs found in \cite{bcz04b} there is no 2-space and these constraints give the relations between simple roots which were required for a type I defect to be momentum conserving. For the Tzitz\'{e}ica defect there is no 1-space and so the constraints obviously hold. These constraints can also be checked to hold for all defects and choices of 1-space and 2-space found in \cite{bb17}, including the $D_4$ defect given in more detail here. Whilst we have not proved anything definite the fact that these constraints have held for all previous momentum conserving defects certainly gives a possible direction for future calculations of defects in $E$ series ATFTs.

As mentioned it is difficult to progress further without any knowledge of the generators appearing in the zero curvature condition, so we will now use these results to show that the momentum conserving Tzitz\'{e}ica and $D_4$ defects given in section \ref{sec:mcd atft} have zero curvature.

\subsection{Zero curvature for the Tzitz\'{e}ica defect} \label{sec:zcc tzitz}

The roots for Tzitz\'{e}ica are given in eq.\eqref{eq:Tzitzroots}, the momentum conserving ATFT defect based on these roots in eq.\eqref{eq:Tzitzdefect} and the momentum conserving defect potential in eqs.\eqref{eq:TzitD3}, \eqref{eq:TzitDbar3}. The defect zero curvature conditions in eqs.\eqref{eq:Khat1}, \eqref{eq:Khat2} then become
\begin{align}
2\sqrt{2}\rho \left[\hat{K},E_{\alpha_0}\right] =&
e^{2f}(e^{q}+e^{-q})^2 \left( \hat{K}_{q} -\{\hat{K},H\} +f_q\left[\hat{K},H\right] \right) \nonumber \\
&+e^{2f}(e^{q}+e^{-q})(e^{q}-e^{-q}) \left[\hat{K},H\right] \label{eq:Tzitzzcc1} \\
2\rho \left[\hat{K},E_{\alpha_1}\right] =&
e^{-f} \left( -\hat{K}_{q} +\{\hat{K},H\} -f_q\left[\hat{K},H\right] \right) \label{eq:Tzitzzcc2} \\
\rho e^{-2f} \left(
\hat{K}_{q}
+f_q \left[\hat{K},H\right] \right)
=&
\sqrt{2}\left( e^{-2q}\hat{K}E_{-\alpha_0} -e^{2q} E_{-\alpha_0}\hat{K} \right) \label{eq:Tzitzzcc3} \\
2\rho e^{f} \big(
(e^{q}+e^{-q}) \left( \hat{K}_{q} +f_q\left[\hat{K},H\right] \right)&
+(e^{q}-e^{-q}) \left[\hat{K},H\right] \big) \nonumber \\
=&
-\left( e^{q}\hat{K}E_{-\alpha_1} -e^{-q} E_{-\alpha_1}\hat{K} \right) \label{eq:Tzitzzcc4}
,\end{align}
where eq.\eqref{eq:Khat2} has been split into two equations by equating powers of $\mu$ and $f$ is some arbitrary function which is present due to our freedom to carry out redefinitions of the auxiliary fields.

In order to solve eqs.\eqref{eq:Tzitzzcc1}-\eqref{eq:Tzitzzcc4} we will choose a representation, write down the generator matrices explicitly, then solve the matrix equations entry by entry to find the elements of $\hat{K}$. For notation we will take $e^n_{i,j}$ to denote an $n\times n$ matrix with zeroes everywhere except position $(i,j)$, where the entry is $1$. Our chosen representation is
\begin{align}
H =& \left(e^3_{1,1}-e^3_{3,3}\right) & E_{\alpha_0} =& e^3_{3,1} & E_{\alpha_1} =& \sqrt{2}\left(e^3_{1,2}+e^3_{2,3}\right)
\end{align}
and we recall that $E_{-\alpha}=E_{\alpha}^{\dagger}$.

Using Maple to solve eqs.\eqref{eq:Tzitzzcc1}-\eqref{eq:Tzitzzcc4} as described then gives
\begin{align}
\hat{K} =& \left(\begin{matrix}
1 -\frac{1}{4\sqrt{2}} \rho^{-3} e^{2q} & \frac{1}{2} \rho^{-2} e^{f} e^{q}(e^{q}+e^{-q}) & -\frac{1}{\sqrt{2}} \rho^{-1} e^{2f} (e^{q}+e^{-q})^2 \\
-\frac{1}{\sqrt{2}} \rho^{-1} e^{-f} & 1 -\frac{1}{4\sqrt{2}} \rho^{-3} & \frac{1}{2} \rho^{-2} e^{f} e^{-q}(e^{q}+e^{-q}) \\
\frac{1}{4} \rho^{-2} e^{-2f} & -\frac{1}{\sqrt{2}} \rho^{-1} e^{-f} & 1 -\frac{1}{4\sqrt{2}} \rho^{-3} e^{-2q}
\end{matrix}\right) \label{eq:TzitzKhatsoln}
.\end{align}
This matrix fits into the proposed form of $\hat{K}$ as a finite series in $\rho$. The structure of this matrix is identical to the Tzitz\'{e}ica defect matrix found in \cite{aagz11}. When writing $\hat{K}$ as given in eq.\eqref{eq:TzitzKhatsoln} in terms of the expansion in $\rho$ given in eq.\eqref{eq:Kexpansion} one possible choice is
\begin{align}
k_0 =& \mathbb{1} \nonumber \\
k_1 =& -\frac{1}{\sqrt{2}} e^{2f} (e^{q}+e^{-q})^2 E_{-\alpha_0} -\frac{1}{2} e^{-f} E_{-\alpha_1} \nonumber \\
k_2 =& \frac{1}{2\sqrt{2}} e^{f} (e^{q}+e^{-q}) \left( e^q E_{-\alpha_0}E_{-\alpha_1} +e^{-q} E_{-\alpha_1}E_{-\alpha_0} \right) +\frac{1}{8} e^{-2f} E_{-\alpha_1}E_{-\alpha_1} \nonumber \\
k_3 =& -\frac{1}{8\sqrt{2}} \left( e^{2q} E_{-\alpha_0}E_{-\alpha_1}E_{-\alpha_1} +E_{-\alpha_1}E_{-\alpha_0}E_{-\alpha_1} +e^{-2q} E_{-\alpha_1}E_{-\alpha_1}E_{-\alpha_0} \right) \label{eq:tzitzk3}
.\end{align}
This fits into the grading hypothesised in the previous chapter, with $k_s$ consisting of products of $s$ generators. Because $K$ appears as part of the monodromy matrix we would hope that $\hat{K}$ could be written as an exponential of generators, but so far such a form of eq.\eqref{eq:TzitzKhatsoln} has not been found. This is due to difficulties with the calculation (at least when carried out in Maple) and there is no proof that it is not possible.

The defect transport matrix satisfying eq.\eqref{eq:defectzcc} is given by
\begin{align}
K=e^{-\frac{1}{2}(p+q-2\mu)H}\hat{K}e^{\frac{1}{2}(p-q-2\mu)H}
.\end{align}

One interesting observation is that there is some additional gauge freedom to that already discussed for the bulk Lax pairs and the defect. Applying no transformations to the bulk Lax pair we can take $K\rightarrow e^{g(q)H}Ke^{-g(q)H}$, so $\hat{K}\rightarrow e^{g(q)H}\hat{K}e^{-g(q)H}$, to give
\begin{align}
\hat{K} =& \left(\begin{matrix}
1 -\frac{1}{4\sqrt{2}} \rho^{-3} e^{2q} & \frac{1}{2} \rho^{-2} e^{f+g} e^{q}(e^{q}+e^{-q}) & -\frac{1}{\sqrt{2}} \rho^{-1} e^{2f+2g} (e^{q}+e^{-q})^2 \\
-\frac{1}{\sqrt{2}} \rho^{-1} e^{-f-g} & 1 -\frac{1}{4\sqrt{2}} \rho^{-3} & \frac{1}{2} \rho^{-2} e^{f+g} e^{-q}(e^{q}+e^{-q}) \\
\frac{1}{4} \rho^{-2} e^{-2f-2g} & -\frac{1}{\sqrt{2}} \rho^{-1} e^{-f-g} & 1 -\frac{1}{4\sqrt{2}} \rho^{-3} e^{-2q}
\end{matrix}\right)
.\end{align}
This transformation obviously corresponds to making the field redefinition $\mu\rightarrow\mu+g(q)$, and so the defect matrix for defects with different definitions of the auxiliary fields are linked by this gauge transformation. The transformed matrix will also satisfy the zero curvature condition, but where before we had $f$ in the defect equations of motion we will now have $f+g$.

\subsection{Zero curvature for the $D_4$ ATFT defect} \label{sec:zcc d4}

The roots for $D_4$ are given in eq.\eqref{eq:D4simpleroots} and the momentum conserving defect Lagrangian in eq.\eqref{eq:LD4}. The two possible momentum conserving defect potentials are given in eqs.\eqref{eq:D4Dgeneral1}, \eqref{eq:D4Dbargeneral1} and eqs.\eqref{eq:D4Dgeneral2}, \eqref{eq:D4Dbargeneral2}. Using the first defect potential ($F=D^++\bar{D}^+$) in eqs.\eqref{eq:Khat1}, \eqref{eq:Khat2} gives
\begin{align}
2\rho \left[\hat{K},E_{\alpha_0}\right] =&
e^{f_{q_2}} \left( \hat{K}_{q_2} -\{\hat{K},H_1\} -\{\hat{K},H_2\} \right) \nonumber \\
&+e^{f_{q_2}}f_{q_2q_2} \left[\hat{K},H_2\right]
+e^{f_{q_2}}f_{q_2q_3} \left[\hat{K},H_3\right] \label{eq:D4zcc1} \\
2\rho \left[\hat{K},E_{\alpha_1}\right] =&
e^{f_{q_2}} \left( \hat{K}_{q_2} +\{\hat{K},H_1\} -\{\hat{K},H_2\} \right) \nonumber \\
&+e^{f_{q_2}}f_{q_2q_2} \left[\hat{K},H_2\right]
+e^{f_{q_2}}f_{q_2q_3} \left[\hat{K},H_3\right] \label{eq:D4zcc2} \\
\sqrt{2} \rho \left[\hat{K},E_{\alpha_2}\right] =&
e^{-f_{q_2}+f_{q_3}}(e^{q_2}+e^{-q_2})
\left( -\hat{K}_{q_2} +\hat{K}_{q_3} +\{\hat{K},H_2\} -\{\hat{K},H_3\} \right) \nonumber \\
&+e^{-f_{q_2}+f_{q_3}}\left( \left(-f_{q_2q_2}+f_{q_2q_3}\right)(e^{q_2}+e^{-q_2}) +e^{q_2}-e^{-q_2} \right) \left[\hat{K},H_2\right] \nonumber \\
&+e^{-f_{q_2}+f_{q_3}}\left(-f_{q_2q_3}+f_{q_3q_3}\right)(e^{q_2}+e^{-q_2}) \left[\hat{K},H_3\right] \label{eq:D4zcc3} \\
2\rho \left[\hat{K},E_{\alpha_3}\right] =&
e^{-f_{q_3}}(e^{q_3}+e^{-q_3})\left( -\hat{K}_{q_3} +\{\hat{K},H_3\} -\{\hat{K},H_4\} \right) \nonumber \\
&-e^{-f_{q_3}}f_{q_2q_3}(e^{q_3}+e^{-q_3}) \left[\hat{K},H_2\right] \nonumber \\
&+e^{-f_{q_3}}\left( -f_{q_3q_3}(e^{q_3}+e^{-q_3}) +e^{q_3}-e^{-q_3} \right) \left[\hat{K},H_3\right] \label{eq:D4zcc4} \\
2\rho \left[\hat{K},E_{\alpha_4}\right] =&
e^{-f_{q_3}}(e^{q_3}+e^{-q_3})\left( -\hat{K}_{q_3} +\{\hat{K},H_3\} +\{\hat{K},H_4\} \right) \nonumber \\
&+-e^{-f_{q_3}}f_{q_2q_3}(e^{q_3}+e^{-q_3}) \left[\hat{K},H_2\right] \nonumber \\
&+e^{-f_{q_3}}\left( -f_{q_3q_3}(e^{q_3}+e^{-q_3}) +e^{q_3}-e^{-q_3} \right) \left[\hat{K},H_3\right] \label{eq:D4zcc5}
\end{align}
\begin{align}
&-2\left(
e^{-q_1-q_2} \hat{K}E_{-\alpha_0}
-e^{q_1+q_2} E_{-\alpha_0}\hat{K}
+e^{q_1-q_2} \hat{K}E_{-\alpha_1}
-e^{-q_1+q_2} E_{-\alpha_1}\hat{K} \right) \nonumber \\
&\qquad=
\rho e^{-f_{q_2}}\big(
-(e^{q_1}+e^{-q_1})(e^{q_2}+e^{-q_2}) \hat{K}_{q_2}
+(e^{q_1}-e^{-q_1})(e^{q_2}+e^{-q_2}) \left[\hat{K},H_1\right] \nonumber \\
&\qquad\qquad\qquad+\left(-f_{q_2q_2}(e^{q_1}+e^{-q_1})(e^{q_2}+e^{-q_2})+(e^{q_1}+e^{-q_1})(e^{q_2}-e^{-q_2})\right) \left[\hat{K},H_2\right] \nonumber \\
&\qquad\qquad\qquad-f_{q_2q_3}(e^{q_1}+e^{-q_1})(e^{q_2}+e^{-q_2}) \left[\hat{K},H_3\right] \big) \label{eq:D4zcc6} \\
&-2\sqrt{2}\left(
e^{q_2-q_3}\hat{K}E_{-\alpha_2}
-e^{-q_2+q_3} E_{-\alpha_2}\hat{K} \right) \nonumber \\
&\qquad=
\rho e^{f_{q_2}-f_{q_3}}\big(
(e^{q_3}+e^{-q_3}) \left(\hat{K}_{q_2} -\hat{K}_{q_3}\right)
+\left(f_{q_2q_2}-f_{q_2q_3}\right)(e^{q_3}+e^{-q_3}) \left[\hat{K},H_2\right] \nonumber \\
&\qquad\qquad\qquad+\left(\left(f_{q_2q_3}-f_{q_3q_3}\right)(e^{q_3}+e^{-q_3}) +e^{q_3}-e^{-q_3}\right) \left[\hat{K},H_3\right] \big) \label{eq:D4zcc7} \\
&-2\left(
e^{q_3-q_4} \hat{K}E_{-\alpha_3}
-e^{-q_3+q_4} E_{-\alpha_3}\hat{K}
+e^{q_3+q_4} \hat{K}E_{-\alpha_4}
-e^{-q_3-q_4} E_{-\alpha_4}\hat{K} \right) \nonumber \\
&\qquad=
\rho e^{f_{q_3}}\big(
(e^{q_4}+e^{-q_4}) \hat{K}_{q_3}
+f_{q_2q_3}(e^{q_4}+e^{-q_4}) \left[\hat{K},H_2\right] \nonumber \\
&\qquad\qquad\qquad+f_{q_3q_3}(e^{q_4}+e^{-q_4}) \left[\hat{K},H_3\right]
+(e^{q_4}-e^{-q_4}) \left[\hat{K},H_4\right] \big)\label{eq:D4zcc8}
\end{align}
and using the second defect potential ($F=D^-+\bar{D}^-$) gives
\begin{align}
2\rho \left[\hat{K},E_{\alpha_0}\right] =&
e^{f_{q_2}} \left( -\hat{K}_{q_2} +\{\hat{K},H_1\} +\{\hat{K},H_2\} \right) \nonumber \\
&-e^{f_{q_2}}f_{q_2q_2} \left[\hat{K},H_2\right]
-e^{f_{q_2}}f_{q_2q_3} \left[\hat{K},H_3\right] \label{eq:D4zcc9} \\
2\rho \left[\hat{K},E_{\alpha_1}\right] =&
e^{f_{q_2}} \left( \hat{K}_{q_2} +\{\hat{K},H_1\} -\{\hat{K},H_2\} \right) \nonumber \\
&+e^{f_{q_2}}f_{q_2q_2} \left[\hat{K},H_2\right]
+e^{f_{q_2}}f_{q_2q_3} \left[\hat{K},H_3\right] \label{eq:D4zcc10} \\
\sqrt{2} \rho \left[\hat{K},E_{\alpha_2}\right] =&
e^{-f_{q_2}+f_{q_3}}(e^{q_2}-e^{-q_2})
\left( \hat{K}_{q_2} -\hat{K}_{q_3} -\{\hat{K},H_2\} +\{\hat{K},H_3\} \right) \nonumber \\
&+e^{-f_{q_2}+f_{q_3}}\left( \left(f_{q_2q_2}-f_{q_2q_3}\right)(e^{q_2}-e^{-q_2}) -e^{q_2}-e^{-q_2} \right) \left[\hat{K},H_2\right] \nonumber \\
&+e^{-f_{q_2}+f_{q_3}}\left(f_{q_2q_3}-f_{q_3q_3}\right)(e^{q_2}-e^{-q_2}) \left[\hat{K},H_3\right] \label{eq:D4zcc11} \\
2\rho \left[\hat{K},E_{\alpha_3}\right] =&
e^{-f_{q_3}}(e^{q_3}-e^{-q_3})\left( \hat{K}_{q_3} -\{\hat{K},H_3\} +\{\hat{K},H_4\} \right) \nonumber \\
&+e^{-f_{q_3}}f_{q_2q_3}(e^{q_3}-e^{-q_3}) \left[\hat{K},H_2\right] \nonumber \\
&+e^{-f_{q_3}}\left( f_{q_3q_3}(e^{q_3}-e^{-q_3}) -e^{q_3}-e^{-q_3} \right) \left[\hat{K},H_3\right] \label{eq:D4zcc12} \\
2\rho \left[\hat{K},E_{\alpha_4}\right] =&
e^{-f_{q_3}}(e^{q_3}-e^{-q_3})\left( -\hat{K}_{q_3} +\{\hat{K},H_3\} +\{\hat{K},H_4\} \right) \nonumber \\
&-e^{-f_{q_3}}f_{q_2q_3}(e^{q_3}-e^{-q_3}) \left[\hat{K},H_2\right] \nonumber \\
&+e^{-f_{q_3}}\left( -f_{q_3q_3}(e^{q_3}-e^{-q_3}) +e^{q_3}+e^{-q_3} \right) \left[\hat{K},H_3\right]\label{eq:D4zcc13}
\end{align}
\begin{align}
&-2\left(
e^{-q_1-q_2} \hat{K}E_{-\alpha_0}
-e^{q_1+q_2} E_{-\alpha_0}\hat{K}
+e^{q_1-q_2} \hat{K}E_{-\alpha_1}
-e^{-q_1+q_2} E_{-\alpha_1}\hat{K} \right) \nonumber \\
&\qquad=
\rho e^{-f_{q_2}}\big(
(e^{q_1}-e^{-q_1})(e^{q_2}-e^{-q_2}) \hat{K}_{q_2}
-(e^{q_1}+e^{-q_1})(e^{q_2}-e^{-q_2}) \left[\hat{K},H_1\right] \nonumber \\
&\qquad\qquad\qquad+\left(f_{q_2q_2}(e^{q_1}-e^{-q_1})(e^{q_2}-e^{-q_2})-(e^{q_1}-e^{-q_1})(e^{q_2}+e^{-q_2})\right) \left[\hat{K},H_2\right] \nonumber \\
&\qquad\qquad\qquad+f_{q_2q_3}(e^{q_1}-e^{-q_1})(e^{q_2}-e^{-q_2}) \left[\hat{K},H_3\right] \big) \label{eq:D4zcc14} \\
&-2\sqrt{2}\left(
e^{q_2-q_3}\hat{K}E_{-\alpha_2}
-e^{-q_2+q_3} E_{-\alpha_2}\hat{K} \right) \nonumber \\
&\qquad=
\rho e^{f_{q_2}-f_{q_3}}\big(
(e^{q_3}-e^{-q_3}) \left(\hat{K}_{q_2} -\hat{K}_{q_3}\right)
+\left(f_{q_2q_2}-f_{q_2q_3}\right)(e^{q_3}-e^{-q_3}) \left[\hat{K},H_2\right] \nonumber \\
&\qquad\qquad\qquad+\left(\left(f_{q_2q_3}-f_{q_3q_3}\right)(e^{q_3}-e^{-q_3}) +e^{q_3}+e^{-q_3}\right) \left[\hat{K},H_3\right] \big) \label{eq:D4zcc15} \\
&-2\left(
e^{q_3-q_4} \hat{K}E_{-\alpha_3}
-e^{-q_3+q_4} E_{-\alpha_3}\hat{K}
+e^{q_3+q_4} \hat{K}E_{-\alpha_4}
-e^{-q_3-q_4} E_{-\alpha_4}\hat{K} \right) \nonumber \\
&\qquad=
\rho e^{f_{q_3}}\big(
(e^{q_4}-e^{-q_4}) \hat{K}_{q_3}
+f_{q_2q_3}(e^{q_4}-e^{-q_4}) \left[\hat{K},H_2\right] \nonumber \\
&\qquad\qquad\qquad+f_{q_3q_3}(e^{q_4}-e^{-q_4}) \left[\hat{K},H_3\right]
+(e^{q_4}+e^{-q_4}) \left[\hat{K},H_4\right] \big)\label{eq:D4zcc16}
\end{align}
where in both cases eq.\eqref{eq:Khat2} has been split into three equations by equating powers of $\mu$.

Again in order to solve these matrix equations we must choose a representation of $D_4$. Using the same notation as in the Tzitz\'{e}ica case we take
\begin{align}
H_1 =& e^8_{1,1}-e^8_{2,2}
&
H_2 =& e^8_{3,3}-e^8_{4,4}
&
H_3 =& e^8_{5,5}-e^8_{6,6}
&
H_4 =& e^8_{7,7}-e^8_{8,8} \\
E_{\alpha_1} =& e^8_{1,3}+e^8_{4,2}
&
E_{\alpha_2} =& e^8_{3,5}+e^8_{6,4}
&
E_{\alpha_3} =& e^8_{5,7}+e^8_{8,6}
&
E_{\alpha_4} =& e^8_{5,8}+e^8_{7,6}\nonumber \\
E_{\alpha_0} =& e^8_{2,3}+e^8_{4,1}
.\end{align}
Using this representation and the expansion of $\hat{K}$ in $\rho$ given in eq.\eqref{eq:Kexpansion} we solve the matrix equations \eqref{eq:D4zcc1}-\eqref{eq:D4zcc8} for the first defect potential, given by eqs.\eqref{eq:D4Dgeneral1}, \eqref{eq:D4Dbargeneral1}, to give
\begin{align}
k_0 =& \mathbb{1} \nonumber \\
k_1 =&
-e^{f_{q_2}} \left(E_{-\alpha_0} +E_{-\alpha_1}\right)
-\sqrt{2}e^{-f_{q_2}+f{q_3}}(e^{q_2}+e^{-q_2}) E_{-\alpha_2} \nonumber \\
&-e^{-f{q_3}}(e^{q_3}+e^{-q_3}) \left(E_{-\alpha_3} +E_{-\alpha_4}\right) \nonumber \\
k_2 =&
e^{2f_{q_2}} E_{-\alpha_0}E_{-\alpha_1}
+\sqrt{2}e^{f_{q_3}} \left(e^{q_2}E_{-\alpha_0}E_{-\alpha_2} +e^{-q_2}E_{-\alpha_2}E_{-\alpha_0}\right) \nonumber \\
&+\sqrt{2}e^{f_{q_3}} \left(e^{q_2}E_{-\alpha_1}E_{-\alpha_2} +e^{-q_2}E_{-\alpha_2}E_{-\alpha_1}\right) \nonumber \\
&+\sqrt{2}e^{-f_{q_2}}(e^{q_2}+e^{-q_2}) \left(e^{q_3}E_{-\alpha_2}E_{-\alpha_3} +e^{-q_3}E_{-\alpha_3}E_{-\alpha_2}\right) \nonumber \\
&+\sqrt{2}e^{-f_{q_2}}(e^{q_2}+e^{-q_2}) \left(e^{q_3}E_{-\alpha_2}E_{-\alpha_4} +e^{-q_3}E_{-\alpha_4}E_{-\alpha_2}\right) \nonumber \\
&+e^{-2f_{q_3}}(e^{q_3}+e^{-q_3})^2 E_{-\alpha_3}E_{-\alpha_4}
\nonumber \\
k_3 =&
-\sqrt{2}e^{f_{q_2}+f_{q_3}}\left(e^{q_2}E_{-\alpha_0}E_{-\alpha_1}E_{-\alpha_2} +e^{-q_2}E_{-\alpha_2}E_{-\alpha_0}E_{-\alpha_1}\right) \nonumber \\
&-\sqrt{2}\left(e^{q_2+q_3}E_{-\alpha_0}E_{-\alpha_2}E_{-\alpha_3} +e^{-q_2-q_3}E_{-\alpha_3}E_{-\alpha_2}E_{-\alpha_0}\right) \nonumber \\
&-\sqrt{2}\left(e^{q_2+q_3}E_{-\alpha_0}E_{-\alpha_2}E_{-\alpha_4} +e^{-q_2-q_3}E_{-\alpha_4}E_{-\alpha_2}E_{-\alpha_0}\right) \nonumber \\
&-\sqrt{2}\left(e^{q_2+q_3}E_{-\alpha_1}E_{-\alpha_2}E_{-\alpha_3} +e^{-q_2-q_3}E_{-\alpha_3}E_{-\alpha_2}E_{-\alpha_1}\right) \nonumber \\
&-\sqrt{2}\left(e^{q_2+q_3}E_{-\alpha_1}E_{-\alpha_2}E_{-\alpha_4} +e^{-q_2-q_3}E_{-\alpha_4}E_{-\alpha_2}E_{-\alpha_1}\right) \nonumber \\
&-\sqrt{2}e^{-f_{q_2}-f_{q_3}}(e^{q_2}+e^{-q_2})(e^{q_3}+e^{-q_3}) \left(e^{q_3}E_{-\alpha_2}E_{-\alpha_3}E_{-\alpha_4} +e^{-q_3}E_{-\alpha_3}E_{-\alpha_4}E_{-\alpha_2}\right) \nonumber \\
k_4 =&
2e^{2f_{q_3}}E_{-\alpha_2}E_{-\alpha_0}E_{-\alpha_1}E_{-\alpha_2} +2e^{-2f_{q_2}}(e^{q_2}+e^{-q_2})^2 E_{-\alpha_2}E_{-\alpha_3}E_{-\alpha_4}E_{-\alpha_2} \nonumber \\
&+\sqrt{2}e^{f_{q_2}} \left(e^{q_2+q_3}E_{-\alpha_0}E_{-\alpha_1}E_{-\alpha_2}E_{-\alpha_3} +e^{-q_2-q_3}E_{-\alpha_3}E_{-\alpha_2}E_{-\alpha_0}E_{-\alpha_1}\right) \nonumber \\
&+\sqrt{2}e^{f_{q_2}} \left(e^{q_2+q_3}E_{-\alpha_0}E_{-\alpha_1}E_{-\alpha_2}E_{-\alpha_4} +e^{-q_2-q_3}E_{-\alpha_4}E_{-\alpha_2}E_{-\alpha_0}E_{-\alpha_1}\right) \nonumber \\
&+\sqrt{2}e^{-f_{q_3}}(e^{q_3}+e^{-q_3}) \left(e^{q_2+q_3}E_{-\alpha_0}E_{-\alpha_2}E_{-\alpha_3}E_{-\alpha_4} +e^{-q_2-q_3}E_{-\alpha_3}E_{-\alpha_4}E_{-\alpha_2}E_{-\alpha_0}\right) \nonumber \\
&+\sqrt{2}e^{-f_{q_3}}(e^{q_3}+e^{-q_3}) \left(e^{q_2+q_3}E_{-\alpha_1}E_{-\alpha_2}E_{-\alpha_3}E_{-\alpha_4} +e^{-q_2-q_3}E_{-\alpha_3}E_{-\alpha_4}E_{-\alpha_2}E_{-\alpha_1}\right) \nonumber \\
k_5 =&
-\sqrt{2}e^{f_{q_2}-f_{q_3}}(e^{q3}+e^{-q3}) \left(e^{q_2+q_3}E_{-\alpha_0}E_{-\alpha_1}E_{-\alpha_2}E_{-\alpha_3}E_{-\alpha_4} +e^{-q_2-q_3}E_{-\alpha_3}E_{-\alpha_4}E_{-\alpha_2}E_{-\alpha_0}E_{-\alpha_1}\right) \nonumber \\
&-2e^{f_{q_3}}\left(e^{q_3}E_{-\alpha_2}E_{-\alpha_0}E_{-\alpha_1}E_{-\alpha_2}E_{-\alpha_3} +e^{-q_3}E_{-\alpha_3}E_{-\alpha_2}E_{-\alpha_0}E_{-\alpha_1}E_{-\alpha_2}\right) \nonumber \\
&-2e^{f_{q_3}}\left(e^{q_3}E_{-\alpha_2}E_{-\alpha_0}E_{-\alpha_1}E_{-\alpha_2}E_{-\alpha_4} +e^{-q_3}E_{-\alpha_4}E_{-\alpha_2}E_{-\alpha_0}E_{-\alpha_1}E_{-\alpha_2}\right) \nonumber \\
&-2e^{-f_{q_2}}(e^{q_2}+e^{-q_2}) \left(e^{q_2}E_{-\alpha_0}E_{-\alpha_2}E_{-\alpha_3}E_{-\alpha_4}E_{-\alpha_2} +e^{-q_2}E_{-\alpha_2}E_{-\alpha_3}E_{-\alpha_4}E_{-\alpha_2}E_{-\alpha_0}\right) \nonumber \\
&-2e^{-f_{q_2}}(e^{q_2}+e^{-q_2}) \left(e^{q_2}E_{-\alpha_1}E_{-\alpha_2}E_{-\alpha_3}E_{-\alpha_4}E_{-\alpha_2} +e^{-q_2}E_{-\alpha_2}E_{-\alpha_3}E_{-\alpha_4}E_{-\alpha_2}E_{-\alpha_1}\right) \nonumber \\
k_6 =&
2e^{2q_2}E_{-\alpha_0}E_{-\alpha_1}E_{-\alpha_2}E_{-\alpha_3}E_{-\alpha_4}E_{-\alpha_2} \nonumber \\
&+2e^{-2q_2}E_{-\alpha_2}E_{-\alpha_3}E_{-\alpha_4}E_{-\alpha_2}E_{-\alpha_0}E_{-\alpha_1} \nonumber \\
&+2e^{2q_3}E_{-\alpha_2}E_{-\alpha_0}E_{-\alpha_1}E_{-\alpha_2}E_{-\alpha_3}E_{-\alpha_4} \nonumber \\
&+2e^{-2q_3}E_{-\alpha_3}E_{-\alpha_4}E_{-\alpha_2}E_{-\alpha_0}E_{-\alpha_1}E_{-\alpha_2} \nonumber \\
&+2E_{-\alpha_0}E_{-\alpha_2}E_{-\alpha_3}E_{-\alpha_4}E_{-\alpha_2}E_{-\alpha_0} \nonumber \\
&+2E_{-\alpha_1}E_{-\alpha_2}E_{-\alpha_3}E_{-\alpha_4}E_{-\alpha_2}E_{-\alpha_1} \nonumber \\
&+2E_{-\alpha_3}E_{-\alpha_2}E_{-\alpha_0}E_{-\alpha_1}E_{-\alpha_2}E_{-\alpha_3} \nonumber \\
&+2E_{-\alpha_4}E_{-\alpha_2}E_{-\alpha_0}E_{-\alpha_1}E_{-\alpha_2}E_{-\alpha_4}
\big)
.\end{align}
Solving eqs.\eqref{eq:D4zcc9}-\eqref{eq:D4zcc16} for the second defect potential, given by  eqs.\eqref{eq:D4Dgeneral2}, \eqref{eq:D4Dbargeneral2}, we have
\begin{align}
k_0 =& \mathbb{1} \nonumber \\
k_1 =&
e^{f_{q_2}} \left(E_{-\alpha_0} -E_{-\alpha_1}\right)
+\sqrt{2}e^{-f_{q_2}+f{q_3}}(e^{q_2}-e^{-q_2}) E_{-\alpha_2} \nonumber \\
&+e^{-f{q_3}}(e^{q_3}-e^{-q_3}) \left(E_{-\alpha_3} -E_{-\alpha_4}\right) \nonumber \\
k_2 =&
-e^{2f_{q_2}} E_{-\alpha_0}E_{-\alpha_1}
+\sqrt{2}e^{f_{q_3}} \left(e^{q_2}E_{-\alpha_0}E_{-\alpha_2} -e^{-q_2}E_{-\alpha_2}E_{-\alpha_0}\right) \nonumber \\
&-\sqrt{2}e^{f_{q_3}} \left(e^{q_2}E_{-\alpha_1}E_{-\alpha_2} -e^{-q_2}E_{-\alpha_2}E_{-\alpha_1}\right) \nonumber \\
&+\sqrt{2}e^{-f_{q_2}}(e^{q_2}-e^{-q_2}) \left(e^{q_3}E_{-\alpha_2}E_{-\alpha_3} -e^{-q_3}E_{-\alpha_3}E_{-\alpha_2}\right) \nonumber \\
&-\sqrt{2}e^{-f_{q_2}}(e^{q_2}-e^{-q_2}) \left(e^{q_3}E_{-\alpha_2}E_{-\alpha_4} -e^{-q_3}E_{-\alpha_4}E_{-\alpha_2}\right) \nonumber \\
&-e^{-2f_{q_3}}(e^{q_3}-e^{-q_3})^2 E_{-\alpha_3}E_{-\alpha_4} \nonumber \\
k_3 =&
-\sqrt{2}e^{f_{q_2}+f_{q_3}}\left(e^{q_2}E_{-\alpha_0}E_{-\alpha_1}E_{-\alpha_2} -e^{-q_2}E_{-\alpha_2}E_{-\alpha_0}E_{-\alpha_1}\right) \nonumber \\
&+\sqrt{2}\left(e^{q_2+q_3}E_{-\alpha_0}E_{-\alpha_2}E_{-\alpha_3} +e^{-q_2-q_3}E_{-\alpha_3}E_{-\alpha_2}E_{-\alpha_0}\right) \nonumber \\
&-\sqrt{2}\left(e^{q_2+q_3}E_{-\alpha_0}E_{-\alpha_2}E_{-\alpha_4} +e^{-q_2-q_3}E_{-\alpha_4}E_{-\alpha_2}E_{-\alpha_0}\right) \nonumber \\
&-\sqrt{2}\left(e^{q_2+q_3}E_{-\alpha_1}E_{-\alpha_2}E_{-\alpha_3} +e^{-q_2-q_3}E_{-\alpha_3}E_{-\alpha_2}E_{-\alpha_1}\right) \nonumber \\
&+\sqrt{2}\left(e^{q_2+q_3}E_{-\alpha_1}E_{-\alpha_2}E_{-\alpha_4} +e^{-q_2-q_3}E_{-\alpha_4}E_{-\alpha_2}E_{-\alpha_1}\right) \nonumber \\
&-\sqrt{2}e^{-f_{q_2}-f_{q_3}}(e^{q_2}-e^{-q_2})(e^{q_3}-e^{-q_3}) \left(e^{q_3}E_{-\alpha_2}E_{-\alpha_3}E_{-\alpha_4} -e^{-q_3}E_{-\alpha_3}E_{-\alpha_4}E_{-\alpha_2}\right) \nonumber \\
k_4 =&
2e^{2f_{q_3}}E_{-\alpha_2}E_{-\alpha_0}E_{-\alpha_1}E_{-\alpha_2} +2e^{-2f_{q_2}}(e^{q_2}-e^{-q_2})^2 E_{-\alpha_2}E_{-\alpha_3}E_{-\alpha_4}E_{-\alpha_2} \nonumber \\
&-\sqrt{2}e^{f_{q_2}} \left(e^{q_2+q_3}E_{-\alpha_0}E_{-\alpha_1}E_{-\alpha_2}E_{-\alpha_3} +e^{-q_2-q_3}E_{-\alpha_3}E_{-\alpha_2}E_{-\alpha_0}E_{-\alpha_1}\right) \nonumber \\
&+\sqrt{2}e^{f_{q_2}} \left(e^{q_2+q_3}E_{-\alpha_0}E_{-\alpha_1}E_{-\alpha_2}E_{-\alpha_4} +e^{-q_2-q_3}E_{-\alpha_4}E_{-\alpha_2}E_{-\alpha_0}E_{-\alpha_1}\right) \nonumber \\
&-\sqrt{2}e^{-f_{q_3}}(e^{q_3}-e^{-q_3}) \left(e^{q2+q_3}E_{-\alpha_0}E_{-\alpha_2}E_{-\alpha_3}E_{-\alpha_4} +e^{-q_2-q_3}E_{-\alpha_3}E_{-\alpha_4}E_{-\alpha_2}E_{-\alpha_0}\right) \nonumber \\
&+\sqrt{2}e^{-f_{q_3}}(e^{q_3}-e^{-q_3}) \left(e^{q2+q_3}E_{-\alpha_1}E_{-\alpha_2}E_{-\alpha_3}E_{-\alpha_4} +e^{-q_2-q_3}E_{-\alpha_3}E_{-\alpha_4}E_{-\alpha_2}E_{-\alpha_1}\right) \nonumber \\
k_5 =&
\sqrt{2}e^{f_{q_2}-f_{q_3}}(e^{q3}-e^{-q3}) \left(e^{q_2+q_3}E_{-\alpha_0}E_{-\alpha_1}E_{-\alpha_2}E_{-\alpha_3}E_{-\alpha_4} +e^{-q_2-q_3}E_{-\alpha_3}E_{-\alpha_4}E_{-\alpha_2}E_{-\alpha_0}E_{-\alpha_1}\right) \nonumber \\
&+2e^{f_{q_3}}\left(e^{q_3}E_{-\alpha_2}E_{-\alpha_0}E_{-\alpha_1}E_{-\alpha_2}E_{-\alpha_3} -e^{-q_3}E_{-\alpha_3}E_{-\alpha_2}E_{-\alpha_0}E_{-\alpha_1}E_{-\alpha_2}\right) \nonumber \\
&-2e^{f_{q_3}}\left(e^{q_3}E_{-\alpha_2}E_{-\alpha_0}E_{-\alpha_1}E_{-\alpha_2}E_{-\alpha_4} -e^{-q_3}E_{-\alpha_4}E_{-\alpha_2}E_{-\alpha_0}E_{-\alpha_1}E_{-\alpha_2}\right) \nonumber \\
&+2e^{-f_{q_2}}(e^{q_2}-e^{-q_2}) \left(e^{q_2}E_{-\alpha_0}E_{-\alpha_2}E_{-\alpha_3}E_{-\alpha_4}E_{-\alpha_2} -e^{-q_2}E_{-\alpha_2}E_{-\alpha_3}E_{-\alpha_4}E_{-\alpha_2}E_{-\alpha_0}\right) \nonumber \\
&-2e^{-f_{q_2}}(e^{q_2}-e^{-q_2}) \left(e^{q_2}E_{-\alpha_1}E_{-\alpha_2}E_{-\alpha_3}E_{-\alpha_4}E_{-\alpha_2} -e^{-q_2}E_{-\alpha_2}E_{-\alpha_3}E_{-\alpha_4}E_{-\alpha_2}E_{-\alpha_1}\right) \nonumber \\
k_6 =&
-2e^{2q_2}E_{-\alpha_0}E_{-\alpha_1}E_{-\alpha_2}E_{-\alpha_3}E_{-\alpha_4}E_{-\alpha_2} \nonumber \\
&-2e^{-2q_2}E_{-\alpha_2}E_{-\alpha_3}E_{-\alpha_4}E_{-\alpha_2}E_{-\alpha_0}E_{-\alpha_1} \nonumber \\
&-2e^{2q_3}E_{-\alpha_2}E_{-\alpha_0}E_{-\alpha_1}E_{-\alpha_2}E_{-\alpha_3}E_{-\alpha_4} \nonumber \\
&-2e^{-2q_3}E_{-\alpha_3}E_{-\alpha_4}E_{-\alpha_2}E_{-\alpha_0}E_{-\alpha_1}E_{-\alpha_2} \nonumber \\
&-2E_{-\alpha_0}E_{-\alpha_2}E_{-\alpha_3}E_{-\alpha_4}E_{-\alpha_2}E_{-\alpha_0} \nonumber \\
&-2E_{-\alpha_2}E_{-\alpha_2}E_{-\alpha_3}E_{-\alpha_4}E_{-\alpha_2}E_{-\alpha_1} \nonumber \\
&-2E_{-\alpha_3}E_{-\alpha_2}E_{-\alpha_0}E_{-\alpha_1}E_{-\alpha_2}E_{-\alpha_3} \nonumber \\
&-2E_{-\alpha_4}E_{-\alpha_2}E_{-\alpha_0}E_{-\alpha_1}E_{-\alpha_2}E_{-\alpha_4}
.\end{align}
These solutions also fit into the proposed grading. We have not checked whether the solutions given here and in eq.\eqref{eq:tzitzk3} are representation independent.

The defect transport matrix satisfying eq.\eqref{eq:defectzcc} is given by
\begin{align}
K =& e^{-\frac{1}{2}\left( (p_{1}+q_{1})H_{1} +(p_{2}+q_{2}-2\mu_{2,t})H_{2} +(p_{3}+q_{3}-2\mu_{3,t})H_{3} +(p_{4}+q_{4})H_{4} \right)} \hat{K} \nonumber \\
&e^{\frac{1}{2}\left( (p_{1}-q_{1})H_{1} +(p_{2}-q_{2}-2\mu_{2,t})H_{2} +(p_{3}-q_{3}-2\mu_{3,t})H_{3} +(p_{4}-q_{4})H_{4} \right)}
.\end{align}

Once again we have $K\rightarrow e^{g(q_2,q_3)_{q_2}H_2+g(q_2,q_3)_{q_3}H_3}Ke^{-g(q_2,q_3)_{q_2}H_2-g(q_2,q_3)_{q_3}H_3}$ taking the $K$ matrix from that of the original defect to that of a defect which is the original defect with the auxiliary fields shifted by $\mu_2\rightarrow\mu_2+g(q_2,q_3)_{q_2}$, $\mu_3\rightarrow\mu_3+g(q_2,q_3)_{q_3}$.

The structure of these defect transport matrices is clearer if we write out the matrices in full. To do this we simplify the situation slightly by setting $f=0$, knowing that the above expression could immediately be used to restore the $e^{f_{q_{2,3}}}$ multipliers to their correct terms. We also take $\hat{K}\rightarrow \frac{1}{\sqrt{2}}\hat{K}$, which does not affect whether $K$ satisfies the zero curvature condition in eq.\eqref{eq:defectzcc}. We use $Q_{2,3}^{\pm}$ to denote the brackets $(e^{q_{2,3}}\pm e^{q_{2,3}})$. The defect matrix for the defect with the first defect potential is
\begin{align}
\hat{K} =& \left(\begin{smallmatrix}
\frac{1}{\sqrt{2}} & \frac{\sqrt{2}}{\rho^{6}} & -\frac{\sqrt{2}e^{q_2}Q_2^+}{\rho^{5}} & -\frac{1}{\sqrt{2}\rho} & \frac{e^{q_2+q_3}Q_3^+}{\rho^{4}} & \frac{e^{q_2}}{\rho^{2}} & -\frac{e^{q_2+q_3}}{\rho^{3}} & -\frac{e^{q_2+q_3}}{\rho^{3}} \\
\frac{\sqrt{2}}{\rho^{6}} & \frac{1}{\sqrt{2}} & -\frac{\sqrt{2}e^{q_2}Q_2^+}{\rho^{5}} & -\frac{1}{\sqrt{2}\rho} & \frac{e^{q_2+q_3}Q_3^+}{\rho^{4}} & \frac{e^{q_2}}{\rho^{2}} & -\frac{e^{q_2+q_3}}{\rho^{3}} & -\frac{e^{q_2+q_3}}{\rho^{3}} \\
-\frac{1}{\sqrt{2}\rho} & -\frac{1}{\sqrt{2}\rho} & \frac{1}{\sqrt{2}}+\frac{\sqrt{2}e^{2q_2}}{\rho^{6}} & \frac{1}{\sqrt{2}\rho^{2}} & -\frac{e^{q_2+q_3}Q_3^+}{\rho^{5}} & -\frac{e^{q_2}}{\rho^{3}} & \frac{e^{q_2+q_3}}{\rho^{4}} & \frac{e^{q_2+q_3}}{\rho^{4}} \\
-\frac{\sqrt{2}e^{-q_2}Q_2^+}{\rho^{5}} & -\frac{\sqrt{2}e^{-q_2}Q_2^+}{\rho^{5}} & \frac{\sqrt{2}Q_2^{+2}}{\rho^{4}} & \frac{1}{\sqrt{2}}+\frac{\sqrt{2}e^{-2q_2}}{\rho^{6}} & -\frac{e^{q_3}Q_2^+Q_3^+}{\rho^{3}} & -\frac{Q_2^+}{\rho} & \frac{e^{q_3}Q_2^+}{\rho^{2}} & \frac{e^{q_3}Q_2^+}{\rho^{2}} \\
\frac{e^{-q_2}}{\rho^{2}} & \frac{e^{-q_2}}{\rho^{2}} & -\frac{Q_2^+}{\rho} & -\frac{e^{-q_2}}{\rho^{3}} & \frac{1}{\sqrt{2}}+\frac{\sqrt{2}e^{2q_3}}{\rho^{6}} & \frac{\sqrt{2}}{\rho^{4}} & -\frac{\sqrt{2}e^{q_3}}{\rho^{5}} & -\frac{\sqrt{2}e^{q_3}}{\rho^{5}} \\
\frac{e^{-q_2-q_3}Q_3^+}{\rho^{4}} & \frac{e^{-q_2-q_3}Q_3^+}{\rho^{4}} & -\frac{e^{-q_3}Q_2^+Q_3^+}{\rho^{3}} & -\frac{e^{-q_2-q_3}Q_3^+}{\rho^{5}} & \frac{Q_3^{+2}}{\sqrt{2}\rho^{2}} & \frac{1}{\sqrt{2}}+\frac{\sqrt{2}e^{-2q_3}}{\rho^{6}} & -\frac{Q_3^+}{\sqrt{2}\rho} & -\frac{Q_3^+}{\sqrt{2}\rho} \\
-\frac{e^{-q_2-q_3}}{\rho^{3}} & -\frac{e^{-q_2-q_3}}{\rho^{3}} & \frac{e^{-q_3}Q_2^+}{\rho^{2}} & \frac{e^{-q_2-q_3}}{\rho^{4}} & -\frac{Q_3^+}{\sqrt{2}\rho} & -\frac{\sqrt{2}e^{-q_3}}{\rho^{5}} & \frac{1}{\sqrt{2}} & \frac{\sqrt{2}}{\rho^{6}} \\
-\frac{e^{-q_2-q_3}}{\rho^{3}} & -\frac{e^{-q_2-q_3}}{\rho^{3}} & \frac{e^{-q_3}Q_2^+}{\rho^{2}} & \frac{e^{-q_2-q_3}}{\rho^{4}} & -\frac{Q_3^+}{\sqrt{2}\rho} & -\frac{\sqrt{2}e^{-q_3}}{\rho^{5}} & \frac{\sqrt{2}}{\rho^6} & \frac{1}{\sqrt{2}}
\end{smallmatrix}\right)
\end{align}
and for the second defect potential we have
\begin{align}
\hat{K} =& \left(\begin{smallmatrix}
\frac{1}{\sqrt{2}} & -\frac{\sqrt{2}}{\rho^{6}} & \frac{\sqrt{2}e^{q_2}Q_2^-}{\rho^{5}} & \frac{1}{\sqrt{2}\rho} & -\frac{e^{q_2+q_3}Q_3^-}{\rho^{4}} & \frac{e^{q_2}}{\rho^{2}} & -\frac{e^{q_2+q_3}}{\rho^{3}} & \frac{e^{q_2+q_3}}{\rho^{3}} \\
-\frac{\sqrt{2}}{\rho^{6}} & \frac{1}{\sqrt{2}} & -\frac{\sqrt{2}e^{q_2}Q_2^-}{\rho^{5}} & -\frac{1}{\sqrt{2}\rho} & \frac{e^{q_2+q_3}Q_3^-}{\rho^{4}} & -\frac{e^{q_2}}{\rho^{2}} & \frac{e^{q_2+q_3}}{\rho^{3}} & -\frac{e^{q_2+q_3}}{\rho^{3}} \\
-\frac{1}{\sqrt{2}\rho} & \frac{1}{\sqrt{2}\rho} & 1-\frac{\sqrt{2}e^{2q_2}}{\rho^{6}} & -\frac{1}{\sqrt{2}\rho^{2}} & \frac{e^{q_2+q_3}Q_3^-}{\rho^{5}} & -\frac{e^{q_2}}{\rho^{3}} & \frac{e^{q_2+q_3}}{\rho^{4}} & -\frac{e^{q_2+q_3}}{\rho^{4}} \\
\frac{\sqrt{2}e^{-q_2}Q_2^-}{\rho^{5}} & -\frac{\sqrt{2}e^{-q_2}Q_2^-}{\rho^{5}} & \frac{\sqrt{2}Q_2^{+2}}{\rho^{4}} & \frac{1}{\sqrt{2}}-\frac{\sqrt{2}e^{-2q_2}}{\rho^{6}} & -\frac{e^{q_3}Q_2^-Q_3^-}{\rho^{3}} & \frac{Q_2^-}{\rho} & -\frac{e^{q_3}Q_2^-}{\rho^{2}} & \frac{e^{q_3}Q_2^-}{\rho^{2}} \\
\frac{e^{-q_2}}{\rho^{2}} & -\frac{e^{-q_2}}{\rho^{2}} & \frac{Q_2^-}{\rho} & \frac{e^{-q_2}}{\rho^{3}} & \frac{1}{\sqrt{2}}-\frac{\sqrt{2}e^{2q_3}}{\rho^{6}} & \frac{\sqrt{2}}{\rho^{4}} & -\frac{\sqrt{2}e^{q_3}}{\rho^{5}} & \frac{\sqrt{2}e^{q_3}}{\rho^{5}} \\
\frac{e^{-q_2-q_3}Q_3^-}{\rho^{4}} & -\frac{e^{-q_2-q_3}Q_3^-}{\rho^{4}} & \frac{e^{-q_3}Q_2^-Q_3^-}{\rho^{3}} & \frac{e^{-q_2-q_3}Q_3^-}{\rho^{5}} & -\frac{Q_3^{+2}}{\sqrt{2}\rho^{2}} & \frac{1}{\sqrt{2}}-\frac{\sqrt{2}e^{-2q_3}}{\rho^{6}} & -\frac{Q_3^-}{\sqrt{2}\rho} & \frac{Q_3^-}{\sqrt{2}\rho} \\
-\frac{e^{-q_2-q_3}}{\rho^{3}} & \frac{e^{-q_2-q_3}}{\rho^{3}} & -\frac{e^{-q_3}Q_2^-}{\rho^{2}} & -\frac{e^{-q_2-q_3}}{\rho^{4}} & \frac{Q_3^-}{\sqrt{2}\rho} & -\frac{\sqrt{2}e^{-q_3}}{\rho^{5}} & \frac{1}{\sqrt{2}} & -\frac{\sqrt{2}}{\rho^{6}} \\
\frac{e^{-q_2-q_3}}{\rho^{3}} & -\frac{e^{-q_2-q_3}}{\rho^{3}} & \frac{e^{-q_3}Q_2^-}{\rho^{2}} & \frac{e^{-q_2-q_3}}{\rho^{4}} & -\frac{Q_3^-}{\sqrt{2}\rho} & \frac{\sqrt{2}e^{-q_3}}{\rho^{5}} & -\frac{\sqrt{2}}{\rho^6} & \frac{1}{\sqrt{2}}
\end{smallmatrix}\right)
.\end{align}

With these defect contributions to the Lax pair which give zero curvature if and only if the equations of motion for a momentum conserving $D_4$ defect are satisfied we have made a step towards proving the integrability of the general momentum conserving defects found in \cite{bb17}. In both the Tzitz\'{e}ica and $D_4$ case momentum conservation gave sufficient constraints on the defect for the generation of an infinite number of conserved quantities. It is very likely that in all cases momentum conservation is necessary for integrability.

\section{Conclusions}

In this paper we have made some small additions to the results found in \cite{bb17}, with the more complete $D_4$ defect potential given in section \ref{sec:mcd d4}. The likely constraints on the 1-space and 2-space splitting found in section \ref{sec:zcc atft} may help to further expand the set of momentum conserving defects in ATFTs if they can be applied to the $E$ series root space.

Most importantly we have applied the defect zero curvature condition to the Tzitz\'{e}ica and $D_4$ ATFT defects and found that requiring momentum conservation was both necessary and sufficient for systems containing these defects to have an infinite number of conserved quantities.

While we have successfully shown that two specific defects have zero curvature, and thus an infinite number of conserved quantities, there is still much work to be done on the integrability of defects. It is not clear how the general ATFT defects could be shown to satisfy the zero curvature condition. Beginning by checking whether the Tzitz\'{e}ica and $D_4$ defect matrices found in sections \ref{sec:zcc tzitz}, \ref{sec:zcc d4} are representation independent, it may be useful to attempt to carry out a representation independent calculation of these matrices. Unlike these two specific examples the zero curvature condition for the defect matrix of a general defect in an ATFT cannot be written explicitly as a matrix equation, and so some more general method of solving it will be necessary.

We have also made no attempt to approach these defects from a Hamiltonian perspective, as has been carried out in \cite{ad12a,ad12b,doi15b,doi16}, and have yet to prove that these defects are integrable. It would be interesting to apply the method given in \cite{cz09b} of moving from a Lagrangian to a Hamiltonian picture to these defects.

Finally we have only considered classical integrability in this paper. Quantum defects are well studied, having been introduced in \cite{dms94a,dms94b} and with defects of the type appearing in this paper being investigated in \cite{cz07,cz10,cz11}. The quantum forms of the defects found in \cite{bb17} have not yet been investigated, but once the quantum transmission matrices are known the quantum integrability of these defects could be investigated.

\section*{Acknowledgements}

This work was supported by an STFC studentship.

\bibliographystyle{JHEP}
\bibliography{ConservedQuantities}

\end{document}